\documentclass[aps,prr,twocolumn,superscriptaddress,floatfix,letter]{revtex4}
\usepackage{graphicx}
\usepackage{amssymb}   
\usepackage{amsfonts}
\usepackage{amsmath}
\usepackage{dsfont}
\usepackage{dcolumn}   
\usepackage{ulem}
\usepackage{bm}
\usepackage[mathscr]{eucal}
\usepackage[dvipsnames]{xcolor}
\usepackage[colorlinks, linkcolor=OrangeRed,citecolor=RoyalBlue,urlcolor=NavyBlue]{hyperref}
\usepackage[all]{hypcap} 
\usepackage{mathtools}
\usepackage{wasysym}

\definecolor{mypurple}{rgb}{0.49,0.18,0.56}
\definecolor{mygold}{rgb}{0.93,0.69,0.13}

\begin{document}
\title{Dynamical phase transitions in the two-dimensional transverse-field Ising model}

\author{Tomohiro Hashizume}
\affiliation{Department of Physics and SUPA, University of Strathclyde, Glasgow G4 0NG, United Kingdom}

\author{Ian P.~McCulloch}
\affiliation{School of Mathematics and Physics, The University of Queensland, St.~Lucia, QLD 4072, Australia}

\author{Jad C.~Halimeh}
\affiliation{Max Planck Institute for the Physics of Complex Systems, N\"othnitzer Stra\ss e 38, 01187 Dresden, Germany}
\affiliation{Physics Department, Technical University of Munich, 85747 Garching, Germany}

\begin{abstract}
We investigate two separate notions of dynamical phase transitions in the two-dimensional nearest-neighbor transverse-field Ising model on a square lattice using matrix product states and a new \textit{hybrid} infinite time-evolving block decimation algorithm, where the model is implemented on an infinitely long cylinder with a finite diameter along which periodic boundary conditions are employed. Starting in an ordered initial state, our numerical results suggest that quenches below the dynamical critical point give rise to a ferromagnetic long-time steady state with the Loschmidt return rate exhibiting \textit{anomalous} cusps even when the order parameter never crosses zero. Within the accessible timescales of our numerics, quenches above the dynamical critical point suggest a paramagnetic long-time steady state with the return rate exhibiting \textit{regular} cusps connected to zero crossings of the order parameter. Additionally, our simulations indicate that quenching slightly above the dynamical critical point leads to a coexistence region where both anomalous and regular cusps appear in the return rate. Quenches from the disordered phase further confirm our main conclusions. Our work supports the recent finding that anomalous cusps arise only when local spin excitations are the energetically dominant quasiparticles. Our results are accessible in modern Rydberg experiments.
\end{abstract}
\date{\today}
\maketitle

\section{Introduction}
Criticality is deeply dependent on dimensionality, which comprises one of the three integral constituents of an equilibrium universality class alongside range of interactions and kind of symmetries (equivalently, number of components of the order parameter) \cite{Cardy_book,Sachdev_book,Ma_book}. For example, it was first shown in 1924 by Ernst Ising that the one-dimensional (1D) nearest-neighbor Ising model has no thermal phase transition -- i.e.,~magnetic order can only exist at zero temperature \cite{Ising1925}. Even though Ising incorrectly surmised from this result that his eponymous model would have no thermal phase transition in any dimension, Onsager's exact solution for the two-dimensional (2D) nearest-neighbor Ising model in 1944 \cite{Onsager1944} established the existence of a thermal phase transition at a critical temperature of $2|J|/\log(1+\sqrt{2})$, with $J$ the spin-coupling constant. Both the 1D and 2D nearest-neighbor Ising models have the same $\mathbb{Z} _2$ symmetry and range of interactions, but the difference in spatial dimension leads to fundamentally different physics. In 1958, Landau and Lifshitz proved that long-range order is impossible in 1D systems with short-range interactions \cite{Landau2013}, thus generalizing Ising's original result. Subsequently in 1969, Thouless and Dyson \cite{Thouless1969,Dyson1969} showed that in 1D Ising chains with ferromagnetic power-law interaction profiles $\propto1/r^\alpha$, with $r$ inter-spin distance and $\alpha>0$, long-range order can persist at finite temperature if and only if $\alpha<2$. 

In recent years, the field of dynamical phase transitions (DPT) in quantum many-body physics has witnessed a surge of activity, not least because of significant advancements in ultracold-atom \cite{Levin_book,Yukalov2011,Bloch2008,Greiner2002} and ion-trap \cite{Porras2004,Kim2009,Jurcevic2014} experiments that made possible achieving evolution times long enough to adequately investigate dynamical criticality in such models. Given a Hamiltonian $\hat{H}(h)$ with $h$ an experimentally accessible control parameter, the most common setup has involved preparing the system in its equilibrium thermal state under some initial Hamiltonian $\hat{H}(h=h_\mathrm{i})$, and then abruptly switching the value of $h:h_\mathrm{i}\to h_\mathrm{f}\neq h_\mathrm{i}$. The consequent dynamics due to this \textit{quantum quench} can then host critical phenomena dependent on both $h_\mathrm{i}$ and $h_\mathrm{f}$. One notion of dynamical criticality resembles the Landau paradigm of phase transitions in equilibrium, where nonanalytic or scaling behavior is sought in the dynamics of the order parameter or two-point correlation and response functions \cite{Moeckel2008,Moeckel2010,Sciolla2010,Sciolla2011,Gambassi2011,
Sciolla2013,Maraga2015,Chandran2013,Smacchia2015,
Mori2018,Zhang2017,Chiocchetta2015,Marcuzzi2016,Chiocchetta2017,
Nicklas2015,Halimeh2017b,Halimeh2017,Karl2017}.
We refer to this type of DPT as DPT-I, and it has been investigated in the transverse-field Ising chain with power-law interaction profiles \cite{Halimeh2017}, and the fully connected ($\alpha=0$) transverse-field Ising model at zero \cite{Sciolla2011,Smacchia2015,Homrighausen2017} and finite \cite{Lang2017,Lang2018} temperature.

The second notion of dynamical criticality, DPT-II, rests on an intuitive analogy. Restricting our discussion to zero temperature for simplicity, we quench the ground state $|\psi_\text{i}\rangle$ of $\hat{H}(h_\mathrm{i})$ with $\hat{H}(h_\mathrm{f})$ and construe the overlap $\langle\psi_\text{i}|\exp[-\text{i}\hat{H}(h_\mathrm{f})t]|\psi_\text{i}\rangle$ as a dynamical analog of the equilibrium thermal partition function, where now complexified time $\text{i}t$ stands as the inverse temperature. Consequently, the return rate

\begin{align}\label{eq:RR}
r(t)=-\lim_{N\to\infty}\frac{1}{N}\ln\big|\langle\psi_\text{i}|\text{e}^{-\text{i}\hat{H}(h_\mathrm{f})t}|\psi_\text{i}\rangle\big|^2,
\end{align}
with $N$ the system size, is now a dynamical analog of the thermal free energy. Just as nonanalyticities in the latter denote the existence of a thermal phase transition at a critical temperature, nonanalyticities in the return rate indicate \textit{dynamical quantum phase transitions} at critical evolution times \cite{Heyl2013,Heyl2014,Heyl2015}. In the last five years significant research effort in DPT-II has culminated in various theoretical studies \cite{Heyl2013,Heyl2014,Heyl2015,Andraschko2014,Vajna2014,
Budich2016,Bhattacharya2017,Heyl2017} and experimental realizations \cite{Jurcevic2017,Flaeschner2018}. In the seminal work of Ref.~\cite{Heyl2013}, two dynamical phases were discovered in the nearest-neighbor transverse-field Ising chain (TFIC), which can be exactly solved by a Jordan-Wigner transformation; see Appendix~\ref{sec:JW}. The first, which we refer to as the \textit{trivial} dynamical phase, is for quenches within the same equilibrium phase where no cusps appear in the return rate. This coincides with the order parameter going asymptotically to zero without crossing it \cite{Calabrese2011,Calabrese2012}. The second is the \textit{regular} phase, which occurs for quenches across the critical point and where cusps appear at equally spaced critical times, with each cusp corresponding to a zero crossing of the order parameter. DPT-II was also investigated in higher dimensions such as in the integrable two-dimensional Kitaev honeycomb \cite{Schmitt2015}, two-dimensional Haldane \cite{Bhattacharya2017b,Bhattacharya2017c}, and three-dimensional $O(N)$ \cite{Weidinger2017} models. However, the original picture of two dynamical phases -- one where the return rate is smooth and a second where it is nonanalytic -- persisted. The two-dimensional nearest-neighbor transverse-field Ising model (TFIM) was also considered in exact diagonalization (ED) \cite{Heyl2018,DeNicola2018} and using a stochastic nonequilibrium approach \cite{DeNicola2018}, albeit for a few sites ($4\times4$ and $3\times5$ sites, respectively), which rendered a valid characterization of critical behavior, present inherently in the thermodynamic limit, impractical.

Recently, it was shown that in 1D transverse-field Ising models with certain interaction profiles beyond nearest-neighbor range \cite{Halimeh2017,Zauner2017,Homrighausen2017,Lang2017,Lang2018,Halimeh2018}, a third \textit{anomalous} phase can occur for certain quenches below the dynamical critical point, in which a new kind of cusps appear in the return rate that are not related to any zero crossings of the order parameter. These anomalous cusps occur when the spectrum of the quench Hamiltonian hosts bound domain walls, whereas they are absent when domain walls are freely propagating \cite{Halimeh2018,Defenu2019}. Unlike TFIC, in TFIM local spin-flip excitations are always energetically favorable due to increased dimensionality even when the interactions are still nearest-neighbor. In particular, domain walls in 2D are always energetically unbounded because they scale as the square root of system size. This in principle suggests that the type of DPT-II criticality in TFIM should significantly differ from that of TFIC, especially that it has been shown in 1D that domain-wall coupling (i.e., the dominance of local spin excitations as lowest-lying quasiparticles) is a necessary condition for the appearance of anomalous cusps in the return rate \cite{Defenu2019}.

\begin{figure}[]
	\centering
	\includegraphics [width=0.48\textwidth]{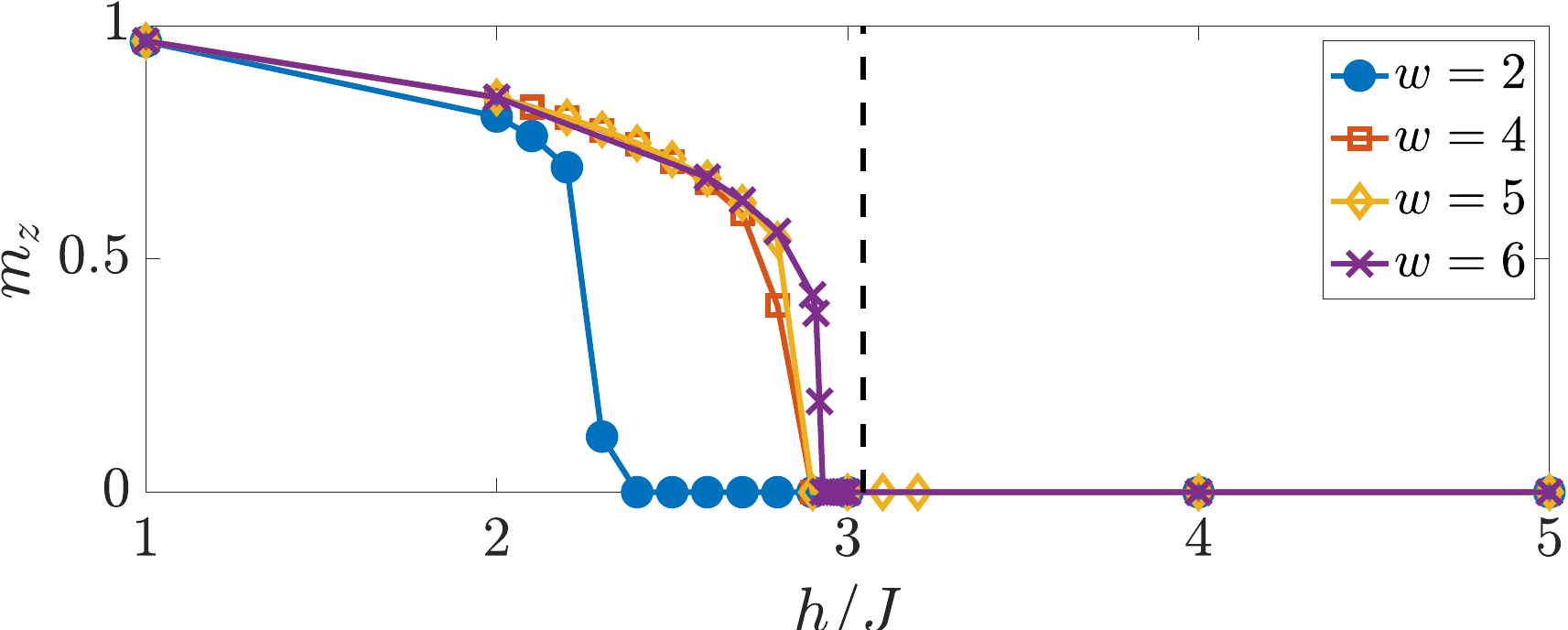}
	\caption{Equilibrium phase diagram of the square-lattice transverse-field Ising model on a cylinder geometry with an infinite-length axis and a $w$-site circumference. The equilibrium critical point obtained from iDMRG is $h_\mathrm{c}^\mathrm{e}\approx2.93J$. The dashed black line indicates the equilibrium quantum critical point of the quantum Ising model on a square lattice in the thermodynamic limit in both directions, as obtained in Refs.~\cite{duCroodeJongh1998,Bloete2002}.}
	\label{fig:EPD}
\end{figure}

In this work, we provide evidence that indeed shows TFIM hosts dynamical criticality that is fundamentally different from that of TFIC, and further validate the conclusions in Ref.~\cite{Halimeh2018} of a quasiparticle origin of the anomalous phase. To the best of our knowledge, our work comprises the first numerically exact study of dynamical phase transitions in nonintegrable higher-dimensional quantum many-body systems in the thermodynamic limit. However, it is important to note here that we achieve the thermodynamic in only one direction, while the second is finite with periodic boundary conditions (see Sec.~\ref{sec:results} for a finite-size analysis). Nevertheless, and as indicated in our results, the sizes we reach in the finite direction approach the thermodynamic limit well both in and out of equilibrium.

\section{Model}
The Hamiltonian of TFIM is

\begin{align}\label{eq:Ham}
\hat{H}(h)=-J\sum_{\langle\mathbf{i},\mathbf{j}\rangle}\hat{\sigma}^z_\mathbf{i}\hat{\sigma}^z_\mathbf{j}-h\sum_\mathbf{j}\hat{\sigma}^x_\mathbf{j},
\end{align}
where $\mathbf{i}$ and $\mathbf{j}$ are lattice vectors, $\langle\mathbf{i},\mathbf{j}\rangle$ indicates nearest-neighbor interactions where each bond is counted only once, and $\hat{\sigma}^{\{x,y,z\}}_\mathbf{j}$ are the Pauli matrices on site $\mathbf{j}$. We build in the framework of the infinite density matrix renormalization group method (iDMRG) \cite{McCulloch2008,IanToolkit} a square lattice on a cylinder geometry of infinite length and a six-site circumference along which periodic boundary conditions are enforced, thereby achieving the thermodynamic limit along the cylinder axis. Let us first consider ferromagnetic interactions ($J>0$) -- as we will see later, this leads to no loss of generality. In the full thermodynamic limit, the square-lattice TFIM has an equilibrium quantum critical point $\approx3.044J$ \cite{duCroodeJongh1998,Bloete2002} and a critical temperature $T_\mathrm{c}=2J/\log(1+\sqrt{2})\approx2.2692J$ \cite{Onsager1944}. However, since in our cylinder geometry the thermodynamic limit is achieved only along the axial direction, finite-size fluctuations due to the six-site circumference lead to a smaller equilibrium quantum critical point $h_\mathrm{c}^\mathrm{e}\approx2.93J$. The equilibrium quantum phase diagram of our model is shown in Fig.~\ref{fig:EPD}, where the ground state is computed through iDMRG.

We have also included results for this phase diagram at smaller values of the number of sites $w$ on the cylinder circumference. As can be seen in Fig.~\ref{fig:EPD}, already at $w=2$ with periodic boundary conditions, the quantum critical point is much larger than that of the nearest-neighbor quantum Ising chain, which is equal to $J$. As $w$ is increased, the quantum critical point approaches that of the quantum Ising square-model in the thermodynamic limit. For the main results of our work, we shall use $w=6$.

\begin{figure}[t!]
	\centering
	\hspace{-.25 cm}
	\includegraphics[width=.49\textwidth]{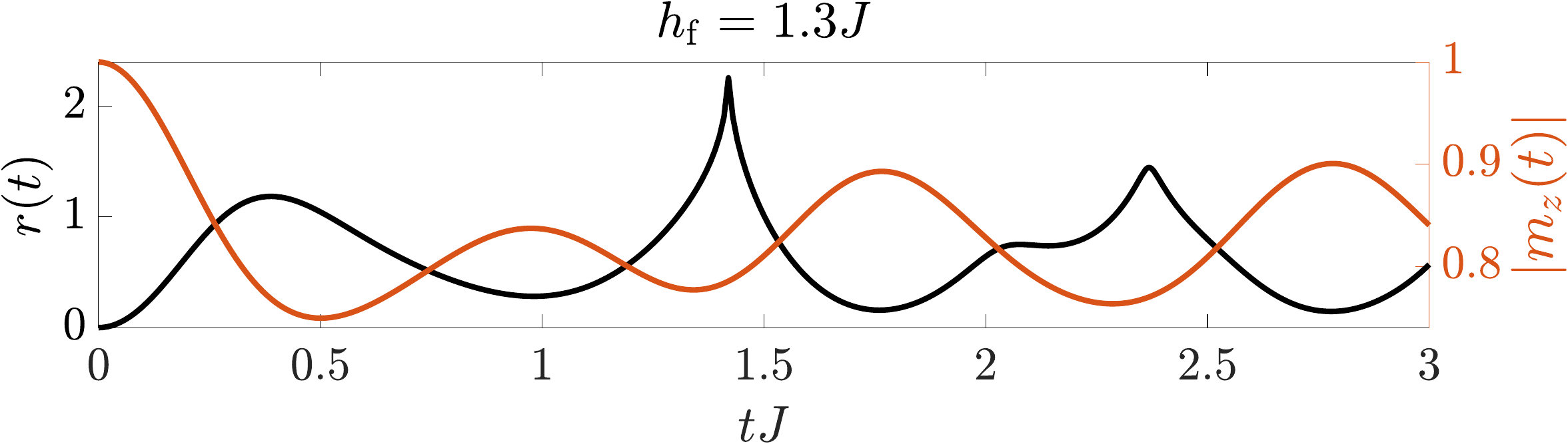}\\
	\hspace{-.25 cm}
	\includegraphics[width=.49\textwidth]{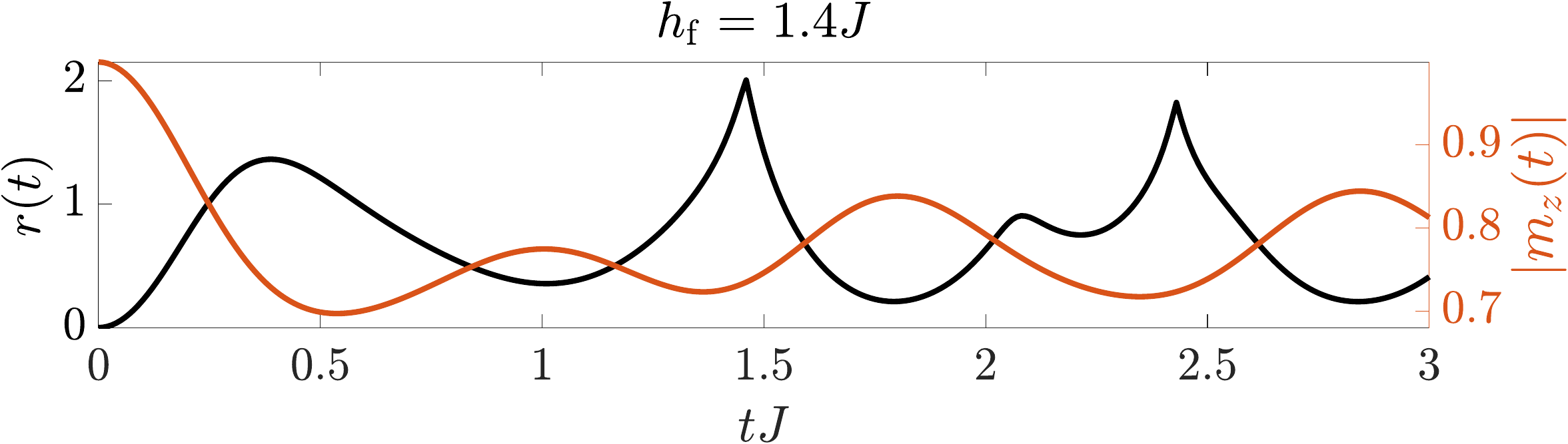}\\
	\hspace{-.25 cm}
	\includegraphics[width=.49\textwidth]{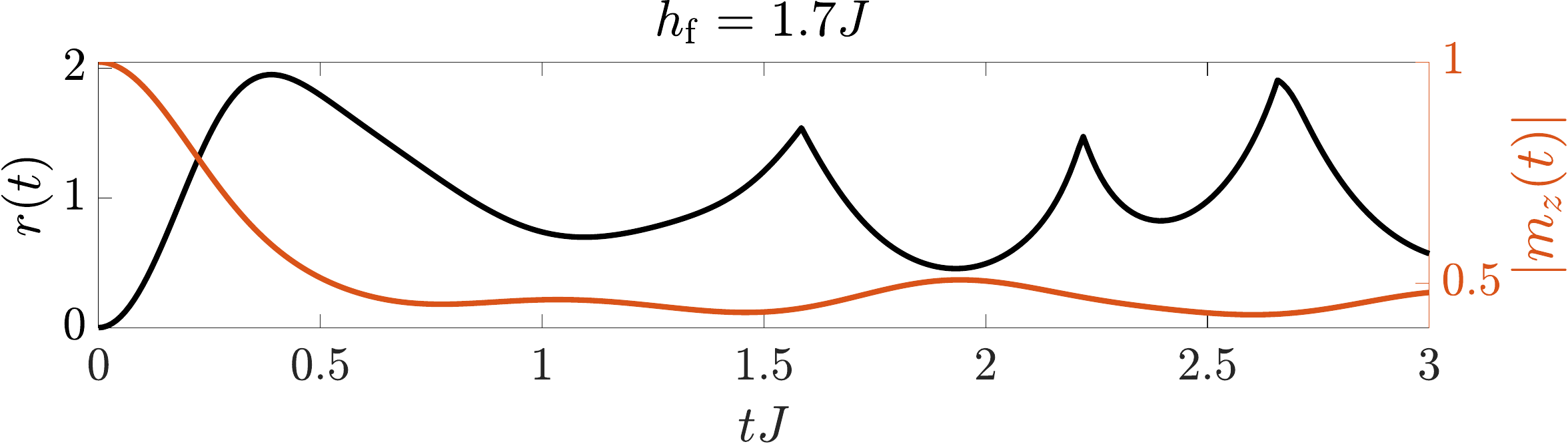}\\
	\hspace{-.25 cm}
	\includegraphics[width=.49\textwidth]{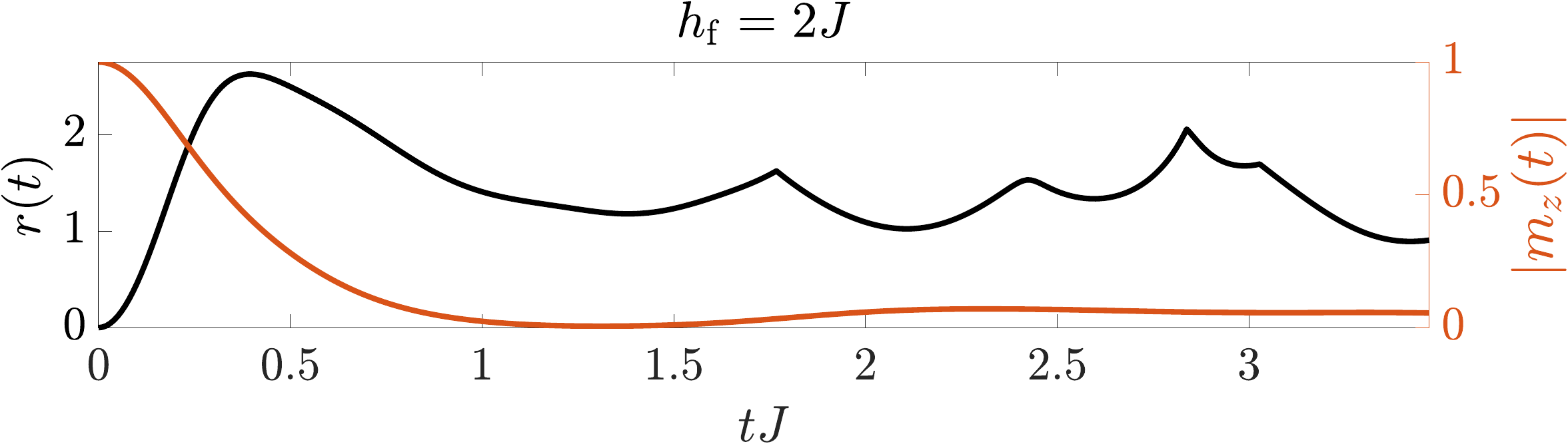}
	\caption{(Color online). Quenches from $h_\text{i}=0$, where the initial state is the fully $z$-up-polarized ground state of TFIM, to final values $h_\text{f}$ of the transverse field below the dynamical critical point $h_\text{c}^\text{d}$. Our results indicate that in this case the order parameter $m_z(t)$ goes asymptotically to a finite nonzero value without ever crossing zero. The return rate always exhibits anomalous cusps.}
	\label{fig:ADP} 
\end{figure}

\section{Results and discussion}\label{sec:results}

We now present our matrix product state (MPS) results for the time evolution of the Loschmidt return rate~\eqref{eq:RR} and the longitudinal and transverse magnetizations

\begin{align}
m_{\{z,x\}}(t)=\lim_{N\to\infty}\frac{1}{N}\sum_\mathbf{j}\langle\hat{\sigma}_\mathbf{j}^{\{z,x\}}(t)\rangle,
\end{align}
respectively, upon quenching the fully ordered ($h_\text{i}=0$) and fully disordered ($h_\text{i}\to\infty$) ground states of TFIM. Our time-evolution results are computed with the hybrid infinite time-evolving block decimation (h-iTEBD) algorithm, the implementation of which can be found in the Matrix Product Toolkit \cite{mptoolkit}. Details on this novel approach are provided in Appendix~\ref{sec:h-iTEBD}, and the full description of its implementation and benchmarking results can be found in Ref.~\cite{Hashizume2019}. Our results reach overall convergence at maximum bond dimension $D_\text{max}=500$ and a time step $\delta t=0.002/J$ (see Appendix~\ref{sec:convergence}).

Let us first consider as initial state the fully $z$-up-polarized ground state ($h_\text{i}=0$) of TFIM. We proceed to quench this state with $\hat{H}(h_\text{f})$, and then calculate the corresponding Loschmidt return rate and order parameter $m_z(t)$. The behavior of the latter critically depends on the value of $h_\text{f}$ to which we quench. Indeed, we find that for quenches below a dynamical critical point $h_\text{c}^\text{d}\approx2.0J$ the order parameter neither crosses nor decays to zero within our accessible evolution times; see Fig.~\ref{fig:ADP}. This behavior is reminiscent of the 1D transverse-field Ising model with power-law ferromagnetic interactions \cite{Halimeh2017,Zauner2017,Homrighausen2017,Lang2017,Lang2018}. For sufficiently long-range interactions ($\alpha<2$), the latter is expected to go into a ferromagnetic steady state in the long-time limit for small quenches due to the model hosting a finite-temperature phase transition. Even when it has no finite-temperature phase transition ($\alpha\geq2$), due to bound domain walls \cite{Liu2018} this system can even settle into a long-lived prethermal state \cite{Halimeh2017b}, which is absent only in the integrable case of nearest-neighbor interactions \cite{Calabrese2011,Calabrese2012} where domain walls freely propagate \cite{Liu2018,Halimeh2018}. Therefore, just like long-range interactions in 1D quantum Ising models give rise to fundamentally different DPT-I criticality, higher dimensionality in the case of TFIM leads to a ferromagnetic steady state for small quenches that does not exist in the case of TFIC. 

\begin{figure}[t!]
	\centering
	\hspace{-.25 cm}
	\includegraphics[width=.49\textwidth]{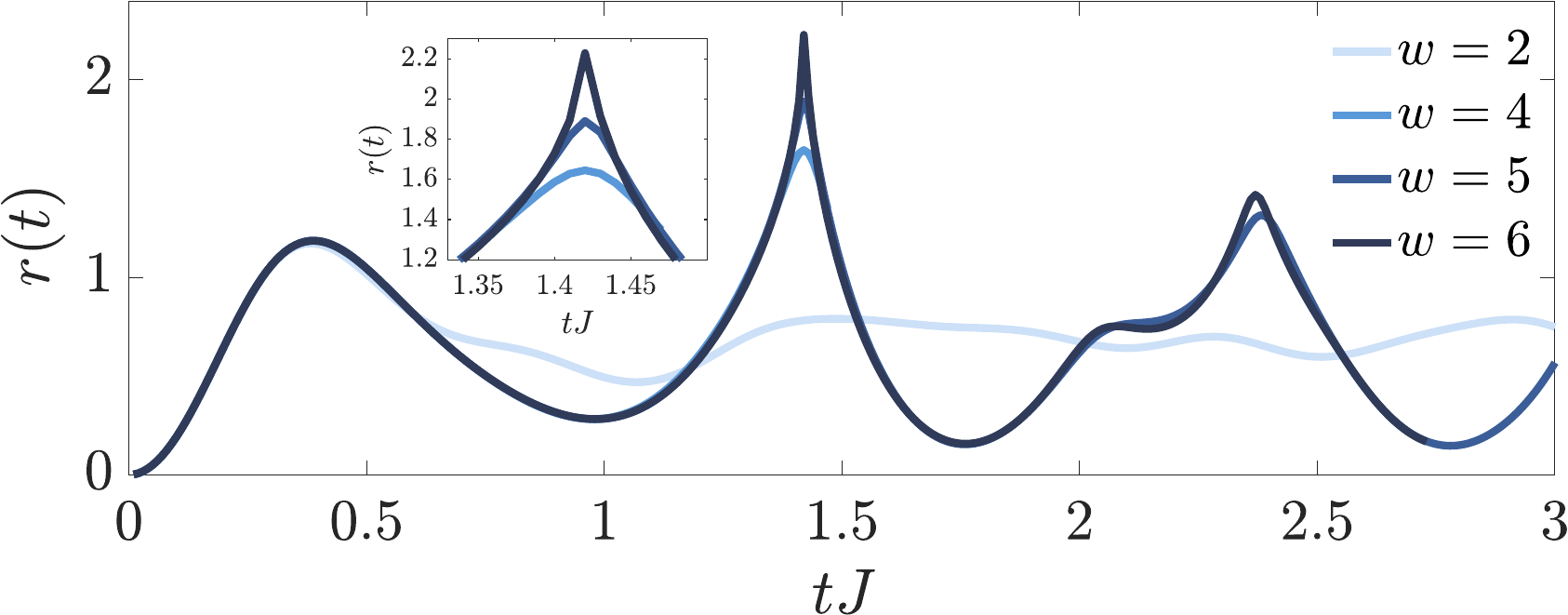}
	\caption{(Color online). Emergence of the anomalous criticality with increasing width $w$ for a quench in the transverse-field strength from $h_\mathrm{i}=0$ to $h_\mathrm{f}=1.3J$, which is well within the ferromagnetic phase for all considered values of $w$. In the limit of a two-leg ladder geometry ($w=2$), we see that the anomalous cusp vanishes. However, as $w$ increases, the return rate exhibits a less smooth behavior, culminating in a clear cusp at $w=6$.}
	\label{fig:FSS} 
\end{figure}

In all cases, $r(t)$ in Fig.~\ref{fig:ADP} exhibits \textit{anomalous} cusps, which, unlike their regular counterparts, are not connected to zero crossings in the order parameter. This resembles return rates due to small quenches in the 1D power-law and exponential-decay interaction models for sufficiently long-range interactions \cite{Halimeh2017,Halimeh2018}. Indeed, we find that the first cycle of the return rate is smooth without any nonanalyticities. This behavior persists even for quenches right below $h_\text{c}^\text{d}$ where the order parameter barely scrapes zero but does not cross it within the timescales of our numerical results; see bottom panel of Fig.~\ref{fig:ADP}. This behavior is fundamentally different from that in TFIC for quenches within the ordered phase where the return rate shows no anomalous cusps and is fully analytic; cf.~Appendix~\ref{sec:JW}. This showcases the crucial effect of dimensionality on DPT-II criticality as well.

Indeed, in two spatial dimensions, single-domain-wall excitations are energetically expensive, and so local spin-flip excitations are the dominant quasiparticles even when interactions are nearest-neighbor. Similarly to the case of long-range interactions in one spatial dimension \cite{Halimeh2018}, when such quasiparticles dominate, anomalous cusps can emerge in the return rate for small quenches within the ferromagnetic phase, as shown in Fig.~\ref{fig:ADP}. These anomalous cusps have no relation to the order parameter changing its sign, in contrast to their regular counterparts \cite{Heyl2013}. 

We further firm up this picture by showing how the anomalous cusp vanishes with decreasing width $w$ of our square lattice along the finite direction. As illustrated in Fig.~\ref{fig:FSS}, we consider the quench in the transverse-field strength from $h_\mathrm{i}=0$ to $h_\mathrm{f}=1.3J$. This quench is well within the ferromagnetic phase of our model for all considered $w$ values. Indeed, for the two-leg ladder geometry ($w=2$), we calculate in iDMRG a critical point $\approx 2.3J$. Similarly to the case in one spatial dimension, we see a smooth return rate at all considered times for $w=2$. However, upon increasing $w$, we see a sharpening of the return rate around the evolution time $t\approx1.42/J$, with a convincing cusp already at $w=5$. This transition in behavior from one typical of one spatial dimension where no cusps arise for quenches below the critical point (small $w$), to one where an anomalous cusp emerges as predicted for a nearest-neighbor interacting two-dimensional system (large $w$) is clear evidence that higher dimensionality and the concomitant dominance of local spin flips in the spectrum of the quench Hamiltonian fundamentally change dynamical criticality. 

\begin{figure}[t!]
\centering
\hspace{-.25 cm}
\includegraphics[width=.49\textwidth]{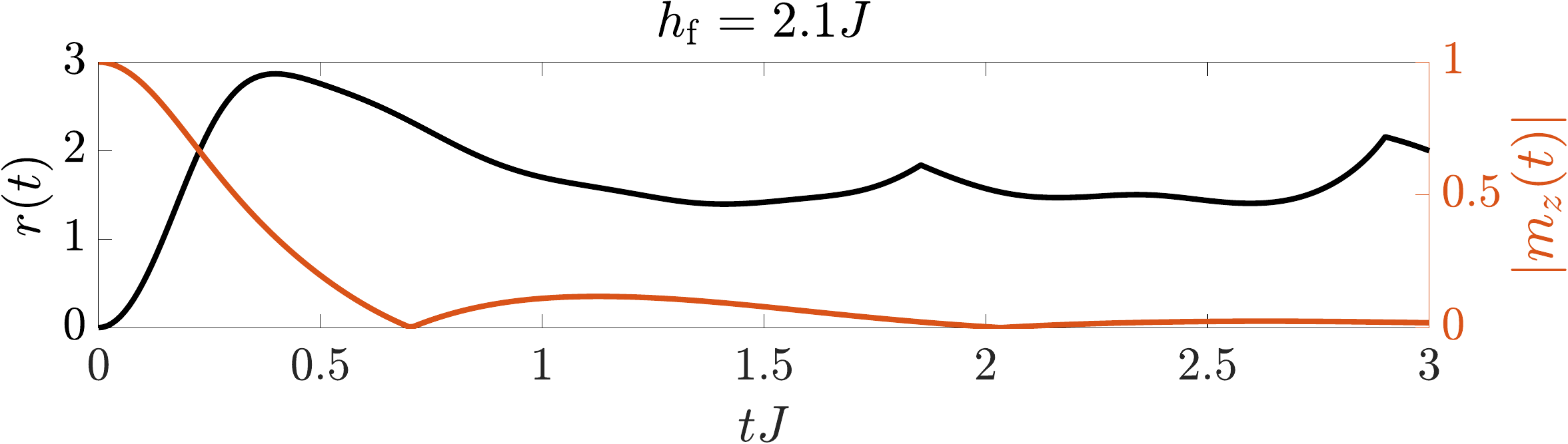}\\
\hspace{-.25 cm}
\includegraphics[width=.49\textwidth]{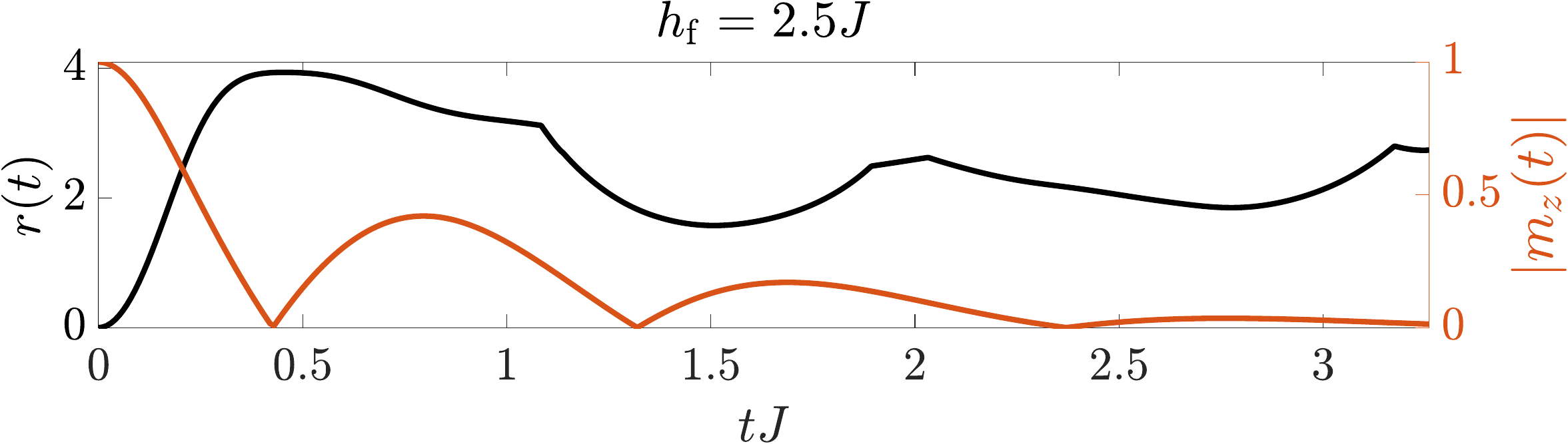}\\
\hspace{-.25 cm}
\includegraphics[width=.49\textwidth]{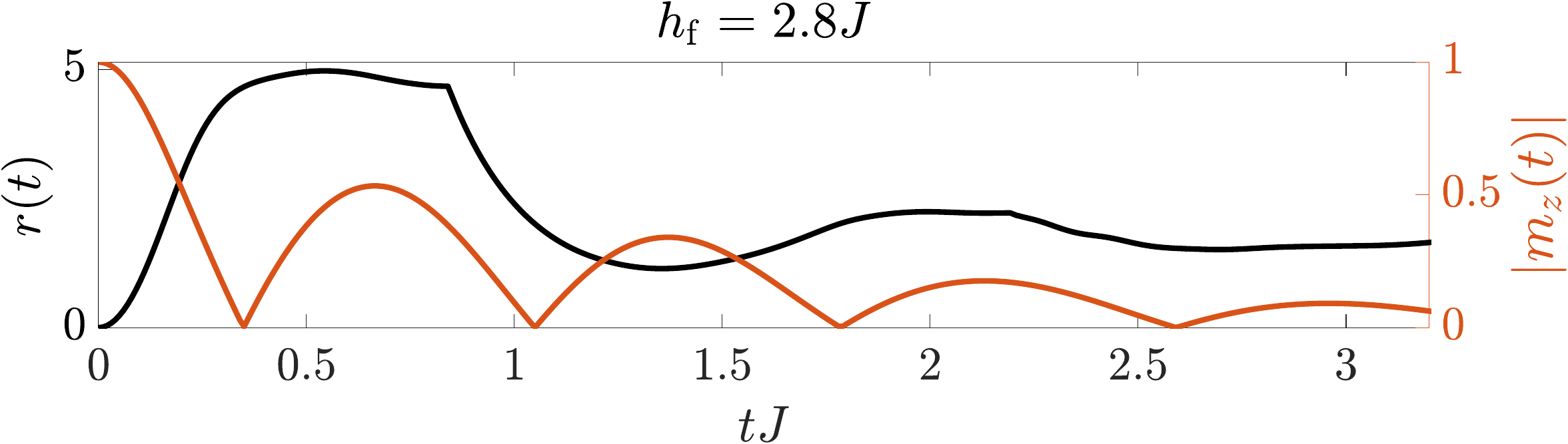}\\
\hspace{-.25 cm}
\includegraphics[width=.49\textwidth]{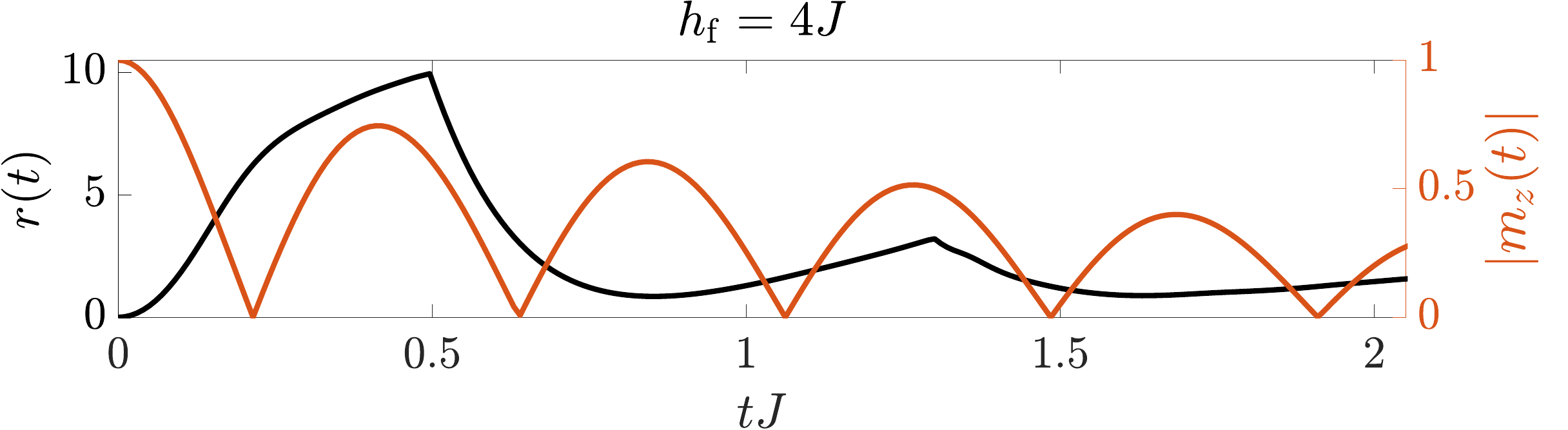}
\caption{(Color online). Same as Fig.~\ref{fig:ADP} but for $h_\text{f}>h_\text{c}^\text{d}$. The order parameter makes zero crosses and seems to asymptotically go to zero. This coincides with the return rate always showing regular cusps, and additionally anomalous cusps when $h_\text{f}\in(h_\text{c}^\text{d},h_\text{cross})$.}
\label{fig:RDP} 
\end{figure}

We now consider quenches to $h_\text{f}>h_\text{c}^\text{d}$ shown in Fig.~\ref{fig:RDP}. Here the order parameter makes zero crossings and its envelope indicates that it goes asymptotically to zero in the long-time limit, as expected for a sufficiently large quench. However, due to our limited timescales in MPS, we cannot ascertain this. The larger $h_\text{f}$ is, the larger the oscillation frequency of the order parameter. At large $h_\text{f}$ (bottom two panels of Fig.~\ref{fig:RDP}), the return rate exhibits a cusp in each cycle such that the periodicity of its cusps is double that of the order-parameter zero crossings, establishing a direct connection between the two in the evolution lifetimes we achieve numerically. We note that this is a mere correlation and one cannot conclude a causal relation. At the smallest value of $h_\text{f}$ (top panel of Fig.~\ref{fig:RDP}) where the order parameter still exhibits zero crossings, the cusps appear to be anomalous rather than regular. This may be because of one of two reasons. The first is that DPT-I and DPT-II may simply not share a common dynamical critical point. The second reason is that there is possibly a coexistence region of both anomalous and regular cusps similar to the case of sufficiently long-range interactions in the 1D case, when the dynamical critical point separating a ferromagnetic steady state from a paramagnetic one is smaller than the crossover value $h_\text{cross}$ of the transverse field below which local spin excitations are energetically dominant \cite{Halimeh2018}. Indeed, in TFIM $h_\text{cross}=h_\text{c}^\text{e}>h_\text{c}^\text{d}$ as domain walls are energetically unbounded in 2D, and thus local spin flips will always be the energetically dominant quasiparticles in the ordered phase. The quench where $h_\text{cross}>h_\text{f}>h_\text{c}^\text{d}$ is exactly when the coexistence region forms for $h_\text{i}=0$ in the dynamical phase diagram of Ref.~\cite{Halimeh2018}. For the existence of this coexistence region to be rigorously confirmed though, we must access in the interval $h_\text{f}/J\in(2,2.3)$ longer evolution times than our code is currently able to achieve in order to discern anomalous from regular cusps. However, lending support to the existence of a coexistance region in the interval $h_\text{f}/J\in(2,2.3)$ is the result in Fig.~\ref{fig:RDP} for $h_\text{f}=2.5J$, where cusps appear at earlier times. There we see the return rate hosting what resembles both regular and anomalous cusps. The first cycle shows a cusp, as is the case in the regular phase, but at the same time the cusps are not evenly spaced in time, which is one of the characteristics of the anomalous phase. This return rate is in great qualitative agreement with those of Ref.~\cite{Halimeh2018} for quenches from $h_\text{i}=0$ to $h_\text{f}\in(h_\text{c}^\text{d},h_\text{cross})$, where the coexistence region has been shown to exist.

The overall picture drawn from the results of Figs.~\ref{fig:ADP} and~\ref{fig:RDP} suggests, therefore, that the anomalous (regular) phase coincides with a ferromagnetic (paramagnetic) long-time steady state, but due to the short evolution times we access in MPS, this cannot be fully ascertained. This is again in remarkable agreement with the cases of the fully connected \cite{Homrighausen2017,Lang2017} and 1D long-range quantum Ising model \cite{Halimeh2017}. It is worth noting here that Refs.~\cite{Heyl2018,DeNicola2018} do not report any anomalous cusps for small quenches within the ordered phase, and we attribute this to the small system sizes they use in their TFIM numerical implementation. Indeed, a large system size is essential to observe anomalous criticality, because the latter is connected to the energetic dominance of local spin flips, while at large system sizes domain walls are energetically unfavorable.

\begin{figure}[t!]
	\centering
	\hspace{-.25 cm}
	\includegraphics[width=.49\textwidth]{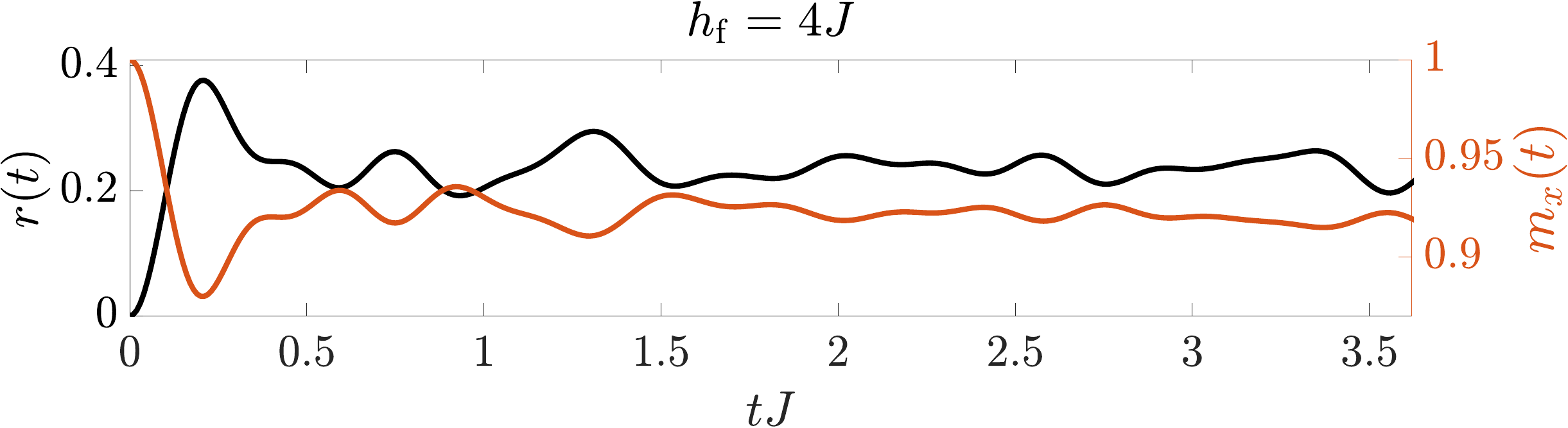}\\
	\hspace{-.25 cm}
	\includegraphics[width=.49\textwidth]{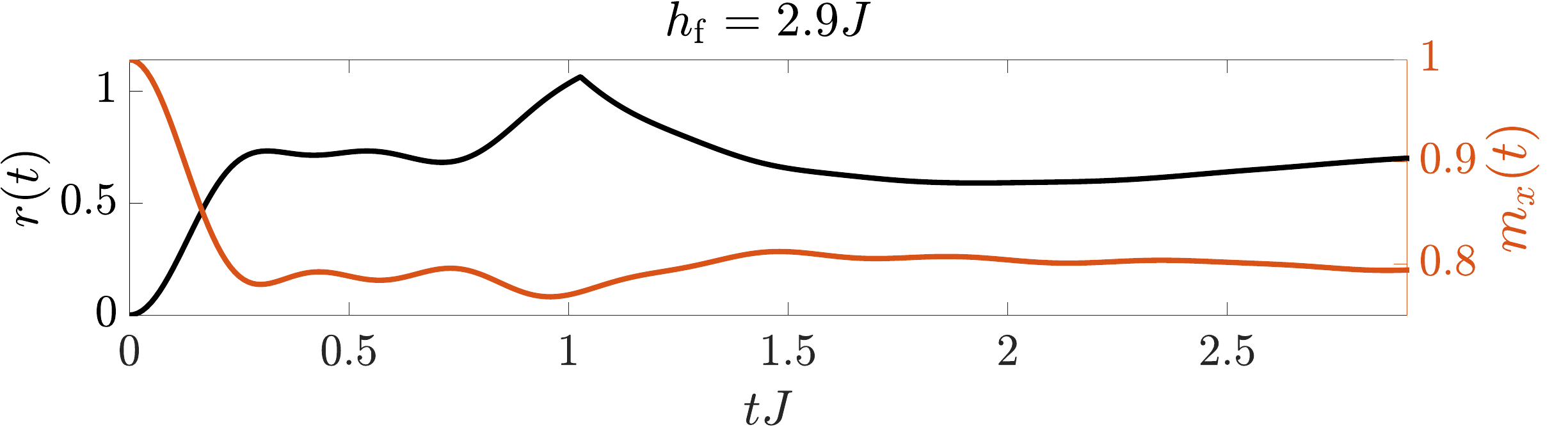}\\
	\hspace{-.25 cm}
	\includegraphics[width=.49\textwidth]{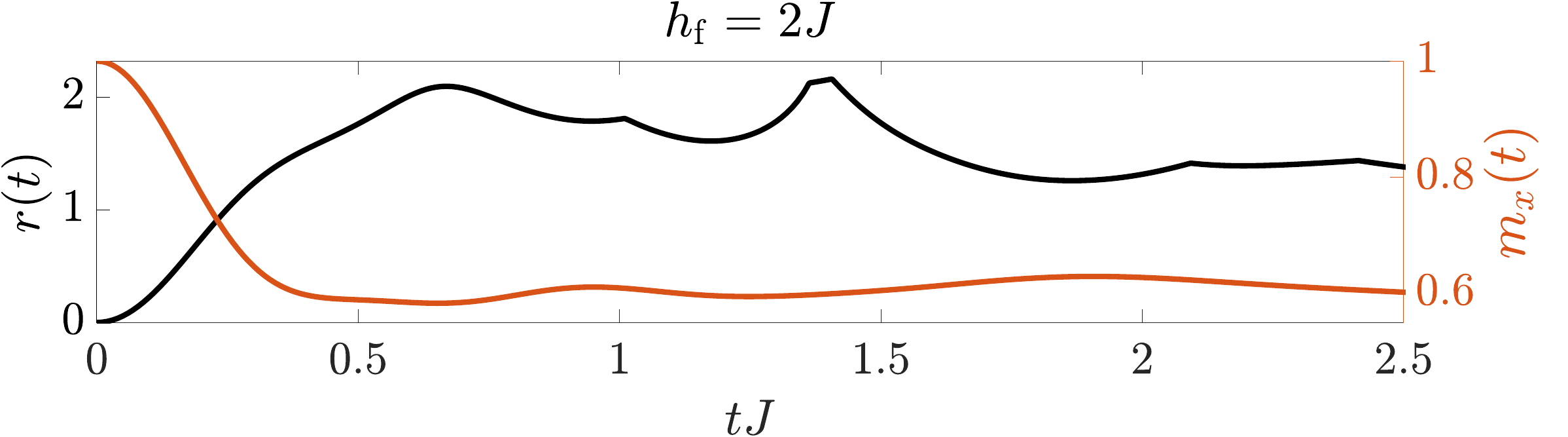}\\
	\hspace{-.25 cm}
	\includegraphics[width=.49\textwidth]{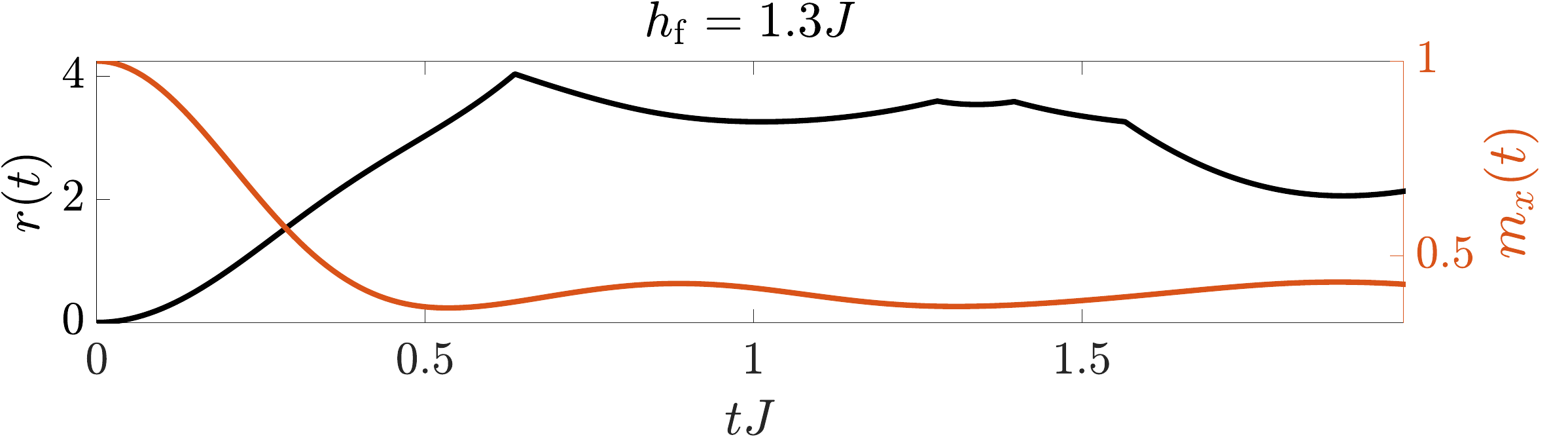}
	\caption{(Color online). Quenches starting from the fully disordered ($h_\text{i}\to\infty$) ground state of TFIM. Small quenches lead to no cusps in the return rate, while those crossing the equilibrium critical point give rise to cusps that are unevenly spaced in time. Since $m_z(t)=0$ at all times, we show instead the transverse magnetization $m_x(t)$.}
	\label{fig:X} 
\end{figure}

We now consider quenches starting in the fully disordered ($h_\text{i}\to\infty$) ground state of TFIM, and quench to various values of $h_\text{f}$. The dynamical critical point in this case is $h_\text{c}^\text{e}$ as in the case of TFIC \cite{Heyl2013}, and is defined based on DPT-II only, since the order parameter is always identically zero. We see two main cases displayed in Fig.~\ref{fig:X}. For quenches within the disordered phase, the return rate shows no cusps, while for quenches to the ordered phase, the return rate displays cusps that are not evenly spaced in time. This is similar to the case of TFIC, except in the latter the cusps always appear at evenly spaced times that are multiples of an analytically determined critical time; cf.~Appendx~\ref{sec:JW}. This is qualitatively identical to what is observed for the same quench in the 1D quantum Ising model with exponentially decaying interactions when domain walls are bound in the spectrum of the quench Hamiltonian, which gives rise to a coexistence region in $r(t)$ \cite{Halimeh2018}.

Finally, we note that the above quenches for the antiferromagnetic case ($J<0$) yield the same behavior qualitatively and quantitatively. This is obvious from the bipartite lattice in a nearest-neighbor model, where ferromagnetic-antiferromagnetic symmetry is up to a spin flip on every second lattice site. This is also the case of TFIC, where the sign of $J$ is inconsequential to the emergent dynamics; see Appendix~\ref{sec:JW} for analytic proof.

\section{Conclusion}We have presented matrix product state results for two notions of dynamical phase transitions in the two-dimensional transverse-field Ising model with nearest-neighbor interactions, showing criticality fundamentally different from the 1D case. When the initial state is ordered, cusps always appear in the return rate regardless of quench distance. Large quenches lead to periodic \textit{regular} cusps with a direct connection to zero crossings of the order parameter, at least within the timescales we achieve in our numerical results. For small quenches, \textit{anomalous} cusps appear that do not show periodicity and are not connected to zero crossings of the order parameter. In a small interval above the dynamical critical point separating a ferromagnetic steady state from a paramagnetic one, our results indicate the formation of a coexistence region in which both anomalous and regular cusps appear in the return rate. This supports results found in Ref.~\cite{Halimeh2018} for the 1D long-range case, where a crossover value of the transverse field -- below which local spin excitations dominate -- is greater than the dynamical critical point, as is the case in TFIM. Moreover, our simulations show within accessible timescales that the anomalous phase overlaps with a ferromagnetic steady state, while the coexistence region and regular phase coincide with a paramagnetic steady state. Quenches from the fully disordered state show no cusps within the disordered phase. On the other hand, when the quench ends in the ordered phase, the return rate shows both regular and anomalous cusps, i.e., the return rate displays the coexistence region, which is found in Ref.~\cite{Halimeh2018} for quenches from the fully disordered state to values of the transverse-field strength below the crossover point.

Our results add credence to the quasiparticle origin of anomalous cusps \cite{Halimeh2018}, are experimentally accessible in modern Rydberg experiments \cite{Zeiher2016,Gross2017}, and usher in the possibility of discerning the long-time steady state properties of a system from the short-time behavior of the return rate.

\section*{Acknowledgments}
J.C.H.~acknowledges stimulating discussions with Bernhard Frank, Christian Gross, Markus Heyl, Johannes Lang, David J.~Luitz, and Daniele Trapin. I.P.M.~acknowledges support from the ARC Future Fellowships scheme, FT140100625.

\appendix
\section{1D nearest-neighbor transverse-field Ising chain}\label{sec:JW}
\noindent The 1D nearest-neighbor transverse-field Ising chain (TFIC) is described by the Hamiltonian

\begin{align}\label{eq:TFIC}
\hat{H}=-\sum_i\big[J\hat{\sigma}^z_i\hat{\sigma}^z_{i+1}+h\hat{\sigma}^x_i\big].
\end{align}
We employ the Jordan-Wigner transformation

\begin{align}
\hat{\sigma}^x_i=&\,1-2\hat{c}_i^\dagger \hat{c}_i,\\
\hat{\sigma}^y_i=&\,-\text{i}\bigg[\prod_{m=1}^{i-1}\big(1-2\hat{c}_m^\dagger \hat{c}_m\big)\bigg]\big(\hat{c}_i-\hat{c}_i^\dagger\big),\\
\hat{\sigma}^z_i=&\,-\bigg[\prod_{m=1}^{i-1}\big(1-2\hat{c}_m^\dagger \hat{c}_m\big)\bigg]\big(\hat{c}_i+\hat{c}_i^\dagger\big),
\end{align}
where $\hat{c}_i,\hat{c}_i^\dagger$ are fermionic annihilation and creation operators, respectively, obeying the canonical anticommutation relations $\{\hat{c}_i,\hat{c}_j\}=0$ and $\{\hat{c}_i,\hat{c}_j^\dagger\}=\delta_{i,j}$. This renders~\eqref{eq:TFIC} in the form

\begin{align}\nonumber
\hat{H}=&\,-\sum_i\big[J\big(\hat{c}_i^\dagger \hat{c}_{i+1}+\hat{c}_i^\dagger \hat{c}_{i+1}^\dagger-\hat{c}_i\hat{c}_{i+1}-\hat{c}_i\hat{c}_{i+1}^\dagger\big)\\\label{eq:TFIC_JW}
&+h\big(1-2\hat{c}_i^\dagger \hat{c}_i\big)\big].
\end{align}
Inserting the Fourier transformation $\hat{c}_i=N^{-1/2}\sum_k^\text{B.z.}\hat{c}_k\mathrm{e}^{\text{i}ki}$, with
$N$ the number of sites, into~\eqref{eq:TFIC_JW}, the Hamiltonian in momentum space takes the form

\begingroup
\renewcommand{\arraystretch}{1.5}
\begin{align}\nonumber
&\hat{H}=\sum_k^\text{B.z.}\hat{\phi}_k^\dagger D_k\hat{\phi}_k,\,\,\,\,\,\,\,\hat{\phi}_k=\,
\begin{pmatrix}
\hat{c}_k \\
\hat{c}_{-k}^\dagger
\end{pmatrix},\\\label{eq:TFIC_JW_FT}
&D_k=\,
\begin{pmatrix}
h-J\cos k & -\text{i}J\sin k \\
\text{i}J\sin k & J\cos k-h
\end{pmatrix}.
\end{align}
\endgroup
The Bogoliubov transformation 

\begingroup
\renewcommand{\arraystretch}{1.5}
\begin{align}
&\hat{\phi}_k=\,M_k\hat{\Gamma}_k,\,\,\,\,\,\,\,M_k=\,
\begin{pmatrix}
\text{i}\sin(\theta_k/2) & \cos(\theta_k/2) \\
\cos(\theta_k/2) & \text{i}\sin(\theta_k/2)
\end{pmatrix},\\\nonumber
&\theta_k=\,\arctan\frac{J\sin k}{h-J\cos k},
\end{align}
\endgroup
diagonalizes~\eqref{eq:TFIC_JW_FT} leading to the dispersion relation

\begin{align}\label{eq:dispersion}
\epsilon_k=\sqrt{(h-J\cos k)^2+J^2\sin^2k}=\sqrt{h^2-2hJ\cos k+J^2}.
\end{align}
Preparing the system in the ground state of the fermionic model in \eqref{eq:TFIC_JW} at an initial value $h_\mathrm{i}$ of the transverse-field strength, and then quenching the system by the same Hamiltonian but at a final value $h_\mathrm{f}$ of the transverse-field strength, we arrive at the return rate \cite{Uhrich2020}
\begin{align}\label{eq:RR-S}
r(t)=&\,-\int_{-\pi}^\pi\frac{\mathrm{d}k}{2\pi}\ln\big[1-\sin^2\big(\theta_k^\mathrm{f}-\theta_k^\mathrm{i}\big)\sin^2\big(2\epsilon_k^\mathrm{f}t\big)\big],
\end{align}
where we have employed the notation $\epsilon_k^{\mathrm{i}(\mathrm{f})}=\epsilon_k(h_{\mathrm{i}(\mathrm{f})})$ and $\theta_k^{\mathrm{i}(\mathrm{f})}=\theta_k(h_{\mathrm{i}(\mathrm{f})})$.
It is therefore clear from~\eqref{eq:RR-S} that nonanalyticities can only occur at critical momenta 

\begin{align}\label{eq:CriticalMomentum}
k_\mathrm{c}=\arccos\frac{J^2+h_\mathrm{i}h_\mathrm{f}}{J(h_\mathrm{i}+h_\mathrm{f})},
\end{align} 
where $|u_{k_\mathrm{c}}|^2=|v_{k_\mathrm{c}}|^2=1/2$, i.e.,~when there is equal probability of occupying both levels in the momentum sector $k_\mathrm{c}$. These nonanalyticities occur at well-specified (periodic) critical times

\begin{align}\label{eq:CriticalTimes}
t_n^*=\bigg(n+\frac{1}{2}\bigg)\frac{\pi}{\epsilon^\text{f}_{k_\mathrm{c}}},\,\,\,\,\,n\in\mathbb{N},
\end{align}
if and only if $h_\mathrm{i}$ and $h_\mathrm{f}$ are on different sides of the equilibrium critical point $h^\text{1D}_\mathrm{c}=|J|$, otherwise $k_\mathrm{c}$, and therefore $t_n^*$, are not well defined.

Already from~\eqref{eq:CriticalTimes} we see a fundamental difference from the case of the two-dimensional transverse-field Ising model (TFIM) discussed in the main text. Whereas here cusps can only occur when crossing a dynamical critical point, which for TFIC coincides with its equilibrium critical point, in the 2D case the cusps occur at any $h_\mathrm{f}\neq h_\mathrm{i}$ as long as $h_\mathrm{i}<h^\text{e}_\mathrm{c}$, the equilibrium critical point of TFIM. The anomalous cusps present for quenches within the ordered phase and below the dynamical critical point in TFIM are due to an underlying quasiparticle spectrum crossover where at small values of the transverse-field strength spin-flip excitations are energetically favorable to two-domain-wall states, as discussed in the main text. This crossover is absent in TFIC, in which two-domain-wall states are always energetically dominant. Moreover,~\eqref{eq:CriticalTimes} indicates a clear periodicity in the return rate after quenches in TFIC, and even though this is also the case for TFIM for quenches deep in the regular phase, in the anomalous phase and coexistence region of the return rate we see cusps that are not evenly spaced in time (see main text). Nevertheless, there is one feature that both models share in that it does not matter whether the interactions are ferromagnetic or antiferromagnetic, the dynamics will be equivalent so long as the interactions are nearest-neighbor. In fact, plugging~\eqref{eq:CriticalMomentum} into~\eqref{eq:dispersion}, it is clear that the sign of $J$ has no effect on the value of $\epsilon_{k_\mathrm{c}}$, which means that the critical times~\eqref{eq:CriticalTimes} are the same for $J=\pm1$.

\section{Hybrid time-evolving block decimation algorithm}\label{sec:h-iTEBD}
In this section, we introduce the hybrid infinite time-evolving block decimation (h-iTEBD) algorithm 
that is used for the time evolution of the states.
h-iTEBD performs a global time evolution 
through the Suzuki-Trotter expansion \cite{Trotter1959,Suzuki1976} and a local time evolution with a method of choice.
Here we choose the Krylov subspace expansion method \cite{Noack2005,Mathematics2015,GarcIa-Ripoll2006} for the local time evolution.

In the ordinary time-evolving block decimation algorithm \cite{Orus2008}, 
a wave function in the thermodynamic limit is described with one pair of $\Gamma$ and $\lambda$ 
matrices (Vidal's notation in Ref.~\cite{Vidal2004}).
Therefore, with this method, only Hamiltonians with nearest-neighbor interactions can be evolved. Although swap gates can be used to force sites to be nearest-neighbor, this is a cumbersome approach that does not readily extend to three- or more-site interactions or exponentially decaying long-range interactions. Here we extend this method so that we can time-evolve Hamiltonians with long-range interactions. By
introducing a unit cell of $L$ sites, with $L$ pairs of $\Gamma_n$ and $\lambda_n$ ($n \in L$), 
we can study systems with dimension greater than one. 

To evolve a state, we first construct two Hamiltonians $\hat{H}_\text{A}$ and $\hat{H}_\text{B}$ such that
$\hat{H}_\text{A}$ describes interactions confined within sites $1$ to $L$ on the unit cell, and $\hat{H}_\text{B}$ describes all interactions 
between site $L/2$ on one unit cell to site $L/2-1$ of the next unit cell with a constraint $\hat{H}=\hat{H}_\text{A}+\hat{H}_\text{B}$, where
$\hat{H}$ is the original Hamiltonian with long-range interactions such that interaction ranges going beyond $L/2$ are truncated. 
By making use of the second-order Suzuki-Trotter formula, we can decompose the infinitesimal time-evolution operator 
$\text{e}^{-\text{i}\delta t\hat{H}}$, with $\delta t\to0$, into a product of local operators that act independently on all of the parts of 
the infinitely long chain:
\begin{align}
\text{e}^{-\text{i}\delta t \hat{H}} \approx 
\text{e}^{-\text{i}\frac{\delta t}{2}\sum_{j}^{\infty} \hat{H}_{\text{A},j}}
\text{e}^{-\text{i}\delta t \sum_{j}^{\infty} \hat{H}_{\text{B},j}}
\text{e}^{-\text{i}\frac{\delta t}{2}\sum_{j}^{\infty} \hat{H}_{\text{A},j}}
+ \mathcal{O}(\delta t^3).
\end{align}
The local time evolution operators $\text{e}^{-\text{i}\frac{\delta t}{2}\sum_{j}^{\infty} \hat{H}_{\text{A},j}}$ and 
$\text{e}^{-\text{i}\delta t \sum_{j}^{\infty} \hat{H}_{\text{B},j}}$ can then be calculated by one's choice of MPS algorithm.
For the calculations that are done in this paper, 
the Krylov subspace expansion algorithm \cite{Noack2005,Mathematics2015,GarcIa-Ripoll2006} 
is used. Due to the leading error of order $\mathcal{O}(\delta t^2)$ from the second-order Suzuki-Trotter expansion,
only three Krylov vectors are calculated for each of 
the local time-evolution operator. This is because with three Krylov vectors the leading error is of the order of 
$\mathcal{O}(\delta t^4)$. The unit cell in the h-iTEBD algorithm can be quite large with no loss of efficiency, which allows for the simulation of long-range interacting models such as the Ising model with power-law decaying interactions \cite{Hashizume_thesis}, and the method applies naturally in finite and infinite settings, as well as infinte boundary conditions \cite{Phien2012}.

The implementation of h-iTEBD is available in the Matrix Product Toolkit \cite{mptoolkit}. The full description and benchmark analysis of h-iTEBD can be found in Ref.~\cite{Hashizume2019}. 

\section{Convergence}\label{sec:convergence}
\begin{figure}[htp]
	\centering
	\hspace{-.25 cm}
	\includegraphics[width=.49\textwidth]{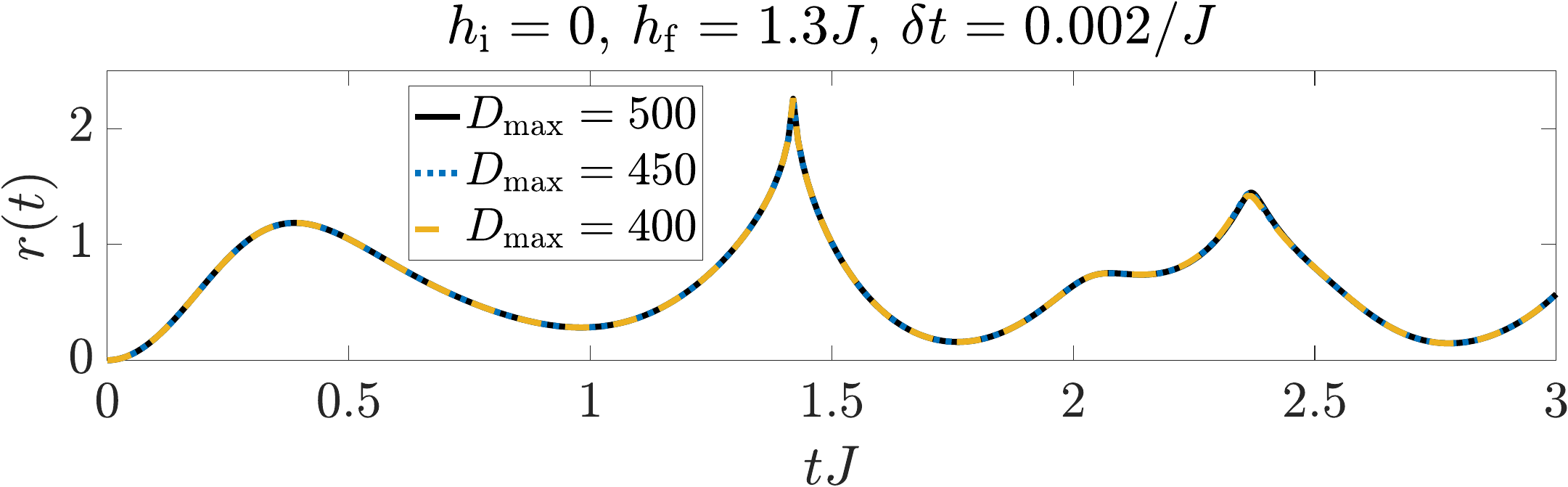}
	\caption{(Color online). For our numerical simulations, we have used various values of the maximum bond dimension $D_\mathrm{max}$. We find convergence at $D_\mathrm{max}=500$ or lower at a time-step of $\delta t=0.002/J$. Here we show a quench from $h_\text{i}=0$ to $h_\text{f}=1.3J$ for illustration.}
	\label{fig:convergence} 
\end{figure}

For our numerical simulations, we find that all results converge at maximum bond dimension $D_\mathrm{max}=500$ and time-step $\delta t=0.002/J$. In Fig.~\ref{fig:convergence}, we show the converged return rate for a quench on the fully $z$-polarized state with $h_\mathrm{f}=1.3J$.

\bibliography{DQPT_biblio}

\begin{thebibliography}{76}
\expandafter\ifx\csname natexlab\endcsname\relax\def\natexlab#1{#1}\fi
\expandafter\ifx\csname bibnamefont\endcsname\relax
  \def\bibnamefont#1{#1}\fi
\expandafter\ifx\csname bibfnamefont\endcsname\relax
  \def\bibfnamefont#1{#1}\fi
\expandafter\ifx\csname citenamefont\endcsname\relax
  \def\citenamefont#1{#1}\fi
\expandafter\ifx\csname url\endcsname\relax
  \def\url#1{\texttt{#1}}\fi
\expandafter\ifx\csname urlprefix\endcsname\relax\def\urlprefix{URL }\fi
\providecommand{\bibinfo}[2]{#2}
\providecommand{\eprint}[2][]{\url{#2}}

\bibitem[{\citenamefont{Cardy}(1996)}]{Cardy_book}
\bibinfo{author}{\bibfnamefont{J.}~\bibnamefont{Cardy}},
  \emph{\bibinfo{title}{Scaling and Renormalization in Statistical Physics}},
  Cambridge Lecture Notes in Physics (\bibinfo{publisher}{Cambridge University
  Press}, \bibinfo{year}{1996}), ISBN \bibinfo{isbn}{9780521499590},
  \urlprefix\url{https://books.google.de/books?id=Wt804S9FjyAC}.

\bibitem[{\citenamefont{Sachdev}(2001)}]{Sachdev_book}
\bibinfo{author}{\bibfnamefont{S.}~\bibnamefont{Sachdev}},
  \emph{\bibinfo{title}{Quantum Phase Transitions}}
  (\bibinfo{publisher}{Cambridge University Press}, \bibinfo{year}{2001}), ISBN
  \bibinfo{isbn}{9780521004541},
  \urlprefix\url{https://books.google.de/books?id=Ih\_E05N5TZQC}.

\bibitem[{\citenamefont{Ma}(1985)}]{Ma_book}
\bibinfo{author}{\bibfnamefont{S.}~\bibnamefont{Ma}},
  \emph{\bibinfo{title}{Statistical Mechanics}} (\bibinfo{publisher}{World
  Scientific}, \bibinfo{year}{1985}), ISBN \bibinfo{isbn}{9789971966065},
  \urlprefix\url{https://books.google.de/books?id=YW-0AQAACAAJ}.

\bibitem[{\citenamefont{{Ising}}(1925)}]{Ising1925}
\bibinfo{author}{\bibfnamefont{E.}~\bibnamefont{{Ising}}},
  \bibinfo{journal}{Zeitschrift fur Physik} \textbf{\bibinfo{volume}{31}},
  \bibinfo{pages}{253} (\bibinfo{year}{1925}).

\bibitem[{\citenamefont{Onsager}(1944)}]{Onsager1944}
\bibinfo{author}{\bibfnamefont{L.}~\bibnamefont{Onsager}},
  \bibinfo{journal}{Phys. Rev.} \textbf{\bibinfo{volume}{65}},
  \bibinfo{pages}{117} (\bibinfo{year}{1944}),
  \urlprefix\url{https://link.aps.org/doi/10.1103/PhysRev.65.117}.

\bibitem[{\citenamefont{Landau and Lifshitz}(2013)}]{Landau2013}
\bibinfo{author}{\bibfnamefont{L.}~\bibnamefont{Landau}} \bibnamefont{and}
  \bibinfo{author}{\bibfnamefont{E.}~\bibnamefont{Lifshitz}},
  \emph{\bibinfo{title}{Statistical Physics}}, \bibinfo{number}{Bd. 5}
  (\bibinfo{publisher}{Elsevier Science}, \bibinfo{year}{2013}), ISBN
  \bibinfo{isbn}{9780080570464},
  \urlprefix\url{https://books.google.de/books?id=VzgJN-XPTRsC}.

\bibitem[{\citenamefont{Thouless}(1969)}]{Thouless1969}
\bibinfo{author}{\bibfnamefont{D.~J.} \bibnamefont{Thouless}},
  \bibinfo{journal}{Phys. Rev.} \textbf{\bibinfo{volume}{187}},
  \bibinfo{pages}{732} (\bibinfo{year}{1969}),
  \urlprefix\url{https://link.aps.org/doi/10.1103/PhysRev.187.732}.

\bibitem[{\citenamefont{Dyson}(1969)}]{Dyson1969}
\bibinfo{author}{\bibfnamefont{F.~J.} \bibnamefont{Dyson}},
  \bibinfo{journal}{F.J. Commun.Math. Phys.} \textbf{\bibinfo{volume}{12}},
  \bibinfo{pages}{91} (\bibinfo{year}{1969}),
  \urlprefix\url{https://link.springer.com/article/10.1007%2FBF01645907#citeas}.

\bibitem[{\citenamefont{Levin et~al.}(2012)\citenamefont{Levin, Fetter, and
  Stamper-Kurn}}]{Levin_book}
\bibinfo{author}{\bibfnamefont{K.}~\bibnamefont{Levin}},
  \bibinfo{author}{\bibfnamefont{A.}~\bibnamefont{Fetter}}, \bibnamefont{and}
  \bibinfo{author}{\bibfnamefont{D.}~\bibnamefont{Stamper-Kurn}},
  \emph{\bibinfo{title}{Ultracold Bosonic and Fermionic Gases}}, Contemporary
  Concepts of Condensed Matter Science (\bibinfo{publisher}{Elsevier Science},
  \bibinfo{year}{2012}), ISBN \bibinfo{isbn}{9780444538628},
  \urlprefix\url{https://books.google.de/books?id=rLlplpMuX6oC}.

\bibitem[{\citenamefont{Yukalov}(2011)}]{Yukalov2011}
\bibinfo{author}{\bibfnamefont{V.~I.} \bibnamefont{Yukalov}},
  \bibinfo{journal}{Laser Physics Letters} \textbf{\bibinfo{volume}{8}},
  \bibinfo{pages}{485} (\bibinfo{year}{2011}),
  \urlprefix\url{http://stacks.iop.org/1612-202X/8/i=7/a=001}.

\bibitem[{\citenamefont{Bloch et~al.}(2008)\citenamefont{Bloch, Dalibard, and
  Zwerger}}]{Bloch2008}
\bibinfo{author}{\bibfnamefont{I.}~\bibnamefont{Bloch}},
  \bibinfo{author}{\bibfnamefont{J.}~\bibnamefont{Dalibard}}, \bibnamefont{and}
  \bibinfo{author}{\bibfnamefont{W.}~\bibnamefont{Zwerger}},
  \bibinfo{journal}{Rev. Mod. Phys.} \textbf{\bibinfo{volume}{80}},
  \bibinfo{pages}{885} (\bibinfo{year}{2008}),
  \urlprefix\url{https://link.aps.org/doi/10.1103/RevModPhys.80.885}.

\bibitem[{\citenamefont{Greiner et~al.}(2002)\citenamefont{Greiner, Mandel,
  H\"ansch, and Bloch}}]{Greiner2002}
\bibinfo{author}{\bibfnamefont{M.}~\bibnamefont{Greiner}},
  \bibinfo{author}{\bibfnamefont{O.}~\bibnamefont{Mandel}},
  \bibinfo{author}{\bibfnamefont{T.~W.} \bibnamefont{H\"ansch}},
  \bibnamefont{and} \bibinfo{author}{\bibfnamefont{I.}~\bibnamefont{Bloch}},
  \bibinfo{journal}{Nature} \textbf{\bibinfo{volume}{419}},
  \bibinfo{pages}{51–54} (\bibinfo{year}{2002}),
  \urlprefix\url{https://www.nature.com/articles/nature00968}.

\bibitem[{\citenamefont{Porras and Cirac}(2004)}]{Porras2004}
\bibinfo{author}{\bibfnamefont{D.}~\bibnamefont{Porras}} \bibnamefont{and}
  \bibinfo{author}{\bibfnamefont{J.~I.} \bibnamefont{Cirac}},
  \bibinfo{journal}{Phys. Rev. Lett.} \textbf{\bibinfo{volume}{92}},
  \bibinfo{pages}{207901} (\bibinfo{year}{2004}),
  \urlprefix\url{https://link.aps.org/doi/10.1103/PhysRevLett.92.207901}.

\bibitem[{\citenamefont{Kim et~al.}(2009)\citenamefont{Kim, Chang, Islam,
  Korenblit, Duan, and Monroe}}]{Kim2009}
\bibinfo{author}{\bibfnamefont{K.}~\bibnamefont{Kim}},
  \bibinfo{author}{\bibfnamefont{M.-S.} \bibnamefont{Chang}},
  \bibinfo{author}{\bibfnamefont{R.}~\bibnamefont{Islam}},
  \bibinfo{author}{\bibfnamefont{S.}~\bibnamefont{Korenblit}},
  \bibinfo{author}{\bibfnamefont{L.-M.} \bibnamefont{Duan}}, \bibnamefont{and}
  \bibinfo{author}{\bibfnamefont{C.}~\bibnamefont{Monroe}},
  \bibinfo{journal}{Phys. Rev. Lett.} \textbf{\bibinfo{volume}{103}},
  \bibinfo{pages}{120502} (\bibinfo{year}{2009}),
  \urlprefix\url{https://link.aps.org/doi/10.1103/PhysRevLett.103.120502}.

\bibitem[{\citenamefont{Jurcevic et~al.}(2014)\citenamefont{Jurcevic, Lanyon,
  Hauke, Hempel, Zoller, Blatt, and Roos}}]{Jurcevic2014}
\bibinfo{author}{\bibfnamefont{P.}~\bibnamefont{Jurcevic}},
  \bibinfo{author}{\bibfnamefont{B.~P.} \bibnamefont{Lanyon}},
  \bibinfo{author}{\bibfnamefont{P.}~\bibnamefont{Hauke}},
  \bibinfo{author}{\bibfnamefont{C.}~\bibnamefont{Hempel}},
  \bibinfo{author}{\bibfnamefont{P.}~\bibnamefont{Zoller}},
  \bibinfo{author}{\bibfnamefont{R.}~\bibnamefont{Blatt}}, \bibnamefont{and}
  \bibinfo{author}{\bibfnamefont{C.~F.} \bibnamefont{Roos}},
  \bibinfo{journal}{Nature} \textbf{\bibinfo{volume}{511}},
  \bibinfo{pages}{202–205} (\bibinfo{year}{2014}),
  \urlprefix\url{http://www.nature.com/articles/nature13461}.

\bibitem[{\citenamefont{Moeckel and Kehrein}(2008)}]{Moeckel2008}
\bibinfo{author}{\bibfnamefont{M.}~\bibnamefont{Moeckel}} \bibnamefont{and}
  \bibinfo{author}{\bibfnamefont{S.}~\bibnamefont{Kehrein}},
  \bibinfo{journal}{Phys. Rev. Lett.} \textbf{\bibinfo{volume}{100}},
  \bibinfo{pages}{175702} (\bibinfo{year}{2008}),
  \urlprefix\url{https://link.aps.org/doi/10.1103/PhysRevLett.100.175702}.

\bibitem[{\citenamefont{Moeckel and Kehrein}(2010)}]{Moeckel2010}
\bibinfo{author}{\bibfnamefont{M.}~\bibnamefont{Moeckel}} \bibnamefont{and}
  \bibinfo{author}{\bibfnamefont{S.}~\bibnamefont{Kehrein}},
  \bibinfo{journal}{New Journal of Physics} \textbf{\bibinfo{volume}{12}},
  \bibinfo{pages}{055016} (\bibinfo{year}{2010}),
  \urlprefix\url{http://stacks.iop.org/1367-2630/12/i=5/a=055016}.

\bibitem[{\citenamefont{Sciolla and Biroli}(2010)}]{Sciolla2010}
\bibinfo{author}{\bibfnamefont{B.}~\bibnamefont{Sciolla}} \bibnamefont{and}
  \bibinfo{author}{\bibfnamefont{G.}~\bibnamefont{Biroli}},
  \bibinfo{journal}{Phys. Rev. Lett.} \textbf{\bibinfo{volume}{105}},
  \bibinfo{pages}{220401} (\bibinfo{year}{2010}),
  \urlprefix\url{https://link.aps.org/doi/10.1103/PhysRevLett.105.220401}.

\bibitem[{\citenamefont{Sciolla and Biroli}(2011)}]{Sciolla2011}
\bibinfo{author}{\bibfnamefont{B.}~\bibnamefont{Sciolla}} \bibnamefont{and}
  \bibinfo{author}{\bibfnamefont{G.}~\bibnamefont{Biroli}},
  \bibinfo{journal}{Journal of Statistical Mechanics: Theory and Experiment}
  \textbf{\bibinfo{volume}{2011}}, \bibinfo{pages}{P11003}
  (\bibinfo{year}{2011}),
  \urlprefix\url{http://stacks.iop.org/1742-5468/2011/i=11/a=P11003}.

\bibitem[{\citenamefont{Gambassi and Calabrese}(2011)}]{Gambassi2011}
\bibinfo{author}{\bibfnamefont{A.}~\bibnamefont{Gambassi}} \bibnamefont{and}
  \bibinfo{author}{\bibfnamefont{P.}~\bibnamefont{Calabrese}},
  \bibinfo{journal}{EPL (Europhysics Letters)} \textbf{\bibinfo{volume}{95}},
  \bibinfo{pages}{66007} (\bibinfo{year}{2011}),
  \urlprefix\url{http://stacks.iop.org/0295-5075/95/i=6/a=66007}.

\bibitem[{\citenamefont{Sciolla and Biroli}(2013)}]{Sciolla2013}
\bibinfo{author}{\bibfnamefont{B.}~\bibnamefont{Sciolla}} \bibnamefont{and}
  \bibinfo{author}{\bibfnamefont{G.}~\bibnamefont{Biroli}},
  \bibinfo{journal}{Phys. Rev. B} \textbf{\bibinfo{volume}{88}},
  \bibinfo{pages}{201110} (\bibinfo{year}{2013}),
  \urlprefix\url{https://link.aps.org/doi/10.1103/PhysRevB.88.201110}.

\bibitem[{\citenamefont{Maraga et~al.}(2015)\citenamefont{Maraga, Chiocchetta,
  Mitra, and Gambassi}}]{Maraga2015}
\bibinfo{author}{\bibfnamefont{A.}~\bibnamefont{Maraga}},
  \bibinfo{author}{\bibfnamefont{A.}~\bibnamefont{Chiocchetta}},
  \bibinfo{author}{\bibfnamefont{A.}~\bibnamefont{Mitra}}, \bibnamefont{and}
  \bibinfo{author}{\bibfnamefont{A.}~\bibnamefont{Gambassi}},
  \bibinfo{journal}{Phys. Rev. E} \textbf{\bibinfo{volume}{92}},
  \bibinfo{pages}{042151} (\bibinfo{year}{2015}),
  \urlprefix\url{https://link.aps.org/doi/10.1103/PhysRevE.92.042151}.

\bibitem[{\citenamefont{Chandran et~al.}(2013)\citenamefont{Chandran, Nanduri,
  Gubser, and Sondhi}}]{Chandran2013}
\bibinfo{author}{\bibfnamefont{A.}~\bibnamefont{Chandran}},
  \bibinfo{author}{\bibfnamefont{A.}~\bibnamefont{Nanduri}},
  \bibinfo{author}{\bibfnamefont{S.~S.} \bibnamefont{Gubser}},
  \bibnamefont{and} \bibinfo{author}{\bibfnamefont{S.~L.}
  \bibnamefont{Sondhi}}, \bibinfo{journal}{Phys. Rev. B}
  \textbf{\bibinfo{volume}{88}}, \bibinfo{pages}{024306}
  (\bibinfo{year}{2013}),
  \urlprefix\url{https://link.aps.org/doi/10.1103/PhysRevB.88.024306}.

\bibitem[{\citenamefont{Smacchia et~al.}(2015)\citenamefont{Smacchia, Knap,
  Demler, and Silva}}]{Smacchia2015}
\bibinfo{author}{\bibfnamefont{P.}~\bibnamefont{Smacchia}},
  \bibinfo{author}{\bibfnamefont{M.}~\bibnamefont{Knap}},
  \bibinfo{author}{\bibfnamefont{E.}~\bibnamefont{Demler}}, \bibnamefont{and}
  \bibinfo{author}{\bibfnamefont{A.}~\bibnamefont{Silva}},
  \bibinfo{journal}{Phys. Rev. B} \textbf{\bibinfo{volume}{91}},
  \bibinfo{pages}{205136} (\bibinfo{year}{2015}),
  \urlprefix\url{https://link.aps.org/doi/10.1103/PhysRevB.91.205136}.

\bibitem[{\citenamefont{Mori et~al.}(2018)\citenamefont{Mori, Ikeda, Kaminishi,
  and Ueda}}]{Mori2018}
\bibinfo{author}{\bibfnamefont{T.}~\bibnamefont{Mori}},
  \bibinfo{author}{\bibfnamefont{T.~N.} \bibnamefont{Ikeda}},
  \bibinfo{author}{\bibfnamefont{E.}~\bibnamefont{Kaminishi}},
  \bibnamefont{and} \bibinfo{author}{\bibfnamefont{M.}~\bibnamefont{Ueda}},
  \bibinfo{journal}{Journal of Physics B: Atomic, Molecular and Optical
  Physics} \textbf{\bibinfo{volume}{51}}, \bibinfo{pages}{112001}
  (\bibinfo{year}{2018}),
  \urlprefix\url{http://stacks.iop.org/0953-4075/51/i=11/a=112001}.

\bibitem[{\citenamefont{Zhang et~al.}(2017)\citenamefont{Zhang, Pagano, Hess,
  Kyprianidis, Becker, Kaplan, Gorshkov, Gong, and Monroe}}]{Zhang2017}
\bibinfo{author}{\bibfnamefont{J.}~\bibnamefont{Zhang}},
  \bibinfo{author}{\bibfnamefont{G.}~\bibnamefont{Pagano}},
  \bibinfo{author}{\bibfnamefont{P.~W.} \bibnamefont{Hess}},
  \bibinfo{author}{\bibfnamefont{A.}~\bibnamefont{Kyprianidis}},
  \bibinfo{author}{\bibfnamefont{P.}~\bibnamefont{Becker}},
  \bibinfo{author}{\bibfnamefont{H.}~\bibnamefont{Kaplan}},
  \bibinfo{author}{\bibfnamefont{A.~V.} \bibnamefont{Gorshkov}},
  \bibinfo{author}{\bibfnamefont{Z.-X.} \bibnamefont{Gong}}, \bibnamefont{and}
  \bibinfo{author}{\bibfnamefont{C.}~\bibnamefont{Monroe}},
  \bibinfo{journal}{Nature} \textbf{\bibinfo{volume}{551}},
  \bibinfo{pages}{601} (\bibinfo{year}{2017}),
  \urlprefix\url{https://www.nature.com/articles/nature24654}.

\bibitem[{\citenamefont{Chiocchetta et~al.}(2015)\citenamefont{Chiocchetta,
  Tavora, Gambassi, and Mitra}}]{Chiocchetta2015}
\bibinfo{author}{\bibfnamefont{A.}~\bibnamefont{Chiocchetta}},
  \bibinfo{author}{\bibfnamefont{M.}~\bibnamefont{Tavora}},
  \bibinfo{author}{\bibfnamefont{A.}~\bibnamefont{Gambassi}}, \bibnamefont{and}
  \bibinfo{author}{\bibfnamefont{A.}~\bibnamefont{Mitra}},
  \bibinfo{journal}{Phys. Rev. B} \textbf{\bibinfo{volume}{91}},
  \bibinfo{pages}{220302} (\bibinfo{year}{2015}),
  \urlprefix\url{https://link.aps.org/doi/10.1103/PhysRevB.91.220302}.

\bibitem[{\citenamefont{Marcuzzi et~al.}(2016)\citenamefont{Marcuzzi, Marino,
  Gambassi, and Silva}}]{Marcuzzi2016}
\bibinfo{author}{\bibfnamefont{M.}~\bibnamefont{Marcuzzi}},
  \bibinfo{author}{\bibfnamefont{J.}~\bibnamefont{Marino}},
  \bibinfo{author}{\bibfnamefont{A.}~\bibnamefont{Gambassi}}, \bibnamefont{and}
  \bibinfo{author}{\bibfnamefont{A.}~\bibnamefont{Silva}},
  \bibinfo{journal}{Phys. Rev. B} \textbf{\bibinfo{volume}{94}},
  \bibinfo{pages}{214304} (\bibinfo{year}{2016}),
  \urlprefix\url{https://link.aps.org/doi/10.1103/PhysRevB.94.214304}.

\bibitem[{\citenamefont{Chiocchetta et~al.}(2017)\citenamefont{Chiocchetta,
  Gambassi, Diehl, and Marino}}]{Chiocchetta2017}
\bibinfo{author}{\bibfnamefont{A.}~\bibnamefont{Chiocchetta}},
  \bibinfo{author}{\bibfnamefont{A.}~\bibnamefont{Gambassi}},
  \bibinfo{author}{\bibfnamefont{S.}~\bibnamefont{Diehl}}, \bibnamefont{and}
  \bibinfo{author}{\bibfnamefont{J.}~\bibnamefont{Marino}},
  \bibinfo{journal}{Phys. Rev. Lett.} \textbf{\bibinfo{volume}{118}},
  \bibinfo{pages}{135701} (\bibinfo{year}{2017}),
  \urlprefix\url{https://link.aps.org/doi/10.1103/PhysRevLett.118.135701}.

\bibitem[{\citenamefont{Nicklas et~al.}(2015)\citenamefont{Nicklas, Karl,
  H\"ofer, Johnson, Muessel, Strobel, Tomkovi\ifmmode~\check{c}\else
  \v{c}\fi{}, Gasenzer, and Oberthaler}}]{Nicklas2015}
\bibinfo{author}{\bibfnamefont{E.}~\bibnamefont{Nicklas}},
  \bibinfo{author}{\bibfnamefont{M.}~\bibnamefont{Karl}},
  \bibinfo{author}{\bibfnamefont{M.}~\bibnamefont{H\"ofer}},
  \bibinfo{author}{\bibfnamefont{A.}~\bibnamefont{Johnson}},
  \bibinfo{author}{\bibfnamefont{W.}~\bibnamefont{Muessel}},
  \bibinfo{author}{\bibfnamefont{H.}~\bibnamefont{Strobel}},
  \bibinfo{author}{\bibfnamefont{J.}~\bibnamefont{Tomkovi\ifmmode~\check{c}\else
  \v{c}\fi{}}}, \bibinfo{author}{\bibfnamefont{T.}~\bibnamefont{Gasenzer}},
  \bibnamefont{and} \bibinfo{author}{\bibfnamefont{M.~K.}
  \bibnamefont{Oberthaler}}, \bibinfo{journal}{Phys. Rev. Lett.}
  \textbf{\bibinfo{volume}{115}}, \bibinfo{pages}{245301}
  (\bibinfo{year}{2015}),
  \urlprefix\url{https://link.aps.org/doi/10.1103/PhysRevLett.115.245301}.

\bibitem[{\citenamefont{Halimeh et~al.}(2017)\citenamefont{Halimeh,
  Zauner-Stauber, McCulloch, de~Vega, Schollw\"ock, and
  Kastner}}]{Halimeh2017b}
\bibinfo{author}{\bibfnamefont{J.~C.} \bibnamefont{Halimeh}},
  \bibinfo{author}{\bibfnamefont{V.}~\bibnamefont{Zauner-Stauber}},
  \bibinfo{author}{\bibfnamefont{I.~P.} \bibnamefont{McCulloch}},
  \bibinfo{author}{\bibfnamefont{I.}~\bibnamefont{de~Vega}},
  \bibinfo{author}{\bibfnamefont{U.}~\bibnamefont{Schollw\"ock}},
  \bibnamefont{and} \bibinfo{author}{\bibfnamefont{M.}~\bibnamefont{Kastner}},
  \bibinfo{journal}{Phys. Rev. B} \textbf{\bibinfo{volume}{95}},
  \bibinfo{pages}{024302} (\bibinfo{year}{2017}),
  \urlprefix\url{https://link.aps.org/doi/10.1103/PhysRevB.95.024302}.

\bibitem[{\citenamefont{Halimeh and Zauner-Stauber}(2017)}]{Halimeh2017}
\bibinfo{author}{\bibfnamefont{J.~C.} \bibnamefont{Halimeh}} \bibnamefont{and}
  \bibinfo{author}{\bibfnamefont{V.}~\bibnamefont{Zauner-Stauber}},
  \bibinfo{journal}{Phys. Rev. B} \textbf{\bibinfo{volume}{96}},
  \bibinfo{pages}{134427} (\bibinfo{year}{2017}),
  \urlprefix\url{https://link.aps.org/doi/10.1103/PhysRevB.96.134427}.

\bibitem[{\citenamefont{Karl et~al.}(2017)\citenamefont{Karl, Cakir, Halimeh,
  Oberthaler, Kastner, and Gasenzer}}]{Karl2017}
\bibinfo{author}{\bibfnamefont{M.}~\bibnamefont{Karl}},
  \bibinfo{author}{\bibfnamefont{H.}~\bibnamefont{Cakir}},
  \bibinfo{author}{\bibfnamefont{J.~C.} \bibnamefont{Halimeh}},
  \bibinfo{author}{\bibfnamefont{M.~K.} \bibnamefont{Oberthaler}},
  \bibinfo{author}{\bibfnamefont{M.}~\bibnamefont{Kastner}}, \bibnamefont{and}
  \bibinfo{author}{\bibfnamefont{T.}~\bibnamefont{Gasenzer}},
  \bibinfo{journal}{Phys. Rev. E} \textbf{\bibinfo{volume}{96}},
  \bibinfo{pages}{022110} (\bibinfo{year}{2017}),
  \urlprefix\url{https://link.aps.org/doi/10.1103/PhysRevE.96.022110}.

\bibitem[{\citenamefont{Homrighausen et~al.}(2017)\citenamefont{Homrighausen,
  Abeling, Zauner-Stauber, and Halimeh}}]{Homrighausen2017}
\bibinfo{author}{\bibfnamefont{I.}~\bibnamefont{Homrighausen}},
  \bibinfo{author}{\bibfnamefont{N.~O.} \bibnamefont{Abeling}},
  \bibinfo{author}{\bibfnamefont{V.}~\bibnamefont{Zauner-Stauber}},
  \bibnamefont{and} \bibinfo{author}{\bibfnamefont{J.~C.}
  \bibnamefont{Halimeh}}, \bibinfo{journal}{Phys. Rev. B}
  \textbf{\bibinfo{volume}{96}}, \bibinfo{pages}{104436}
  (\bibinfo{year}{2017}),
  \urlprefix\url{https://link.aps.org/doi/10.1103/PhysRevB.96.104436}.

\bibitem[{\citenamefont{Lang et~al.}(2018{\natexlab{a}})\citenamefont{Lang,
  Frank, and Halimeh}}]{Lang2017}
\bibinfo{author}{\bibfnamefont{J.}~\bibnamefont{Lang}},
  \bibinfo{author}{\bibfnamefont{B.}~\bibnamefont{Frank}}, \bibnamefont{and}
  \bibinfo{author}{\bibfnamefont{J.~C.} \bibnamefont{Halimeh}},
  \bibinfo{journal}{Phys. Rev. B} \textbf{\bibinfo{volume}{97}},
  \bibinfo{pages}{174401} (\bibinfo{year}{2018}{\natexlab{a}}),
  \urlprefix\url{https://link.aps.org/doi/10.1103/PhysRevB.97.174401}.

\bibitem[{\citenamefont{Lang et~al.}(2018{\natexlab{b}})\citenamefont{Lang,
  Frank, and Halimeh}}]{Lang2018}
\bibinfo{author}{\bibfnamefont{J.}~\bibnamefont{Lang}},
  \bibinfo{author}{\bibfnamefont{B.}~\bibnamefont{Frank}}, \bibnamefont{and}
  \bibinfo{author}{\bibfnamefont{J.~C.} \bibnamefont{Halimeh}},
  \bibinfo{journal}{Phys. Rev. Lett.} \textbf{\bibinfo{volume}{121}},
  \bibinfo{pages}{130603} (\bibinfo{year}{2018}{\natexlab{b}}),
  \urlprefix\url{https://link.aps.org/doi/10.1103/PhysRevLett.121.130603}.

\bibitem[{\citenamefont{Heyl et~al.}(2013)\citenamefont{Heyl, Polkovnikov, and
  Kehrein}}]{Heyl2013}
\bibinfo{author}{\bibfnamefont{M.}~\bibnamefont{Heyl}},
  \bibinfo{author}{\bibfnamefont{A.}~\bibnamefont{Polkovnikov}},
  \bibnamefont{and} \bibinfo{author}{\bibfnamefont{S.}~\bibnamefont{Kehrein}},
  \bibinfo{journal}{Phys. Rev. Lett.} \textbf{\bibinfo{volume}{110}},
  \bibinfo{pages}{135704} (\bibinfo{year}{2013}),
  \urlprefix\url{https://link.aps.org/doi/10.1103/PhysRevLett.110.135704}.

\bibitem[{\citenamefont{Heyl}(2014)}]{Heyl2014}
\bibinfo{author}{\bibfnamefont{M.}~\bibnamefont{Heyl}}, \bibinfo{journal}{Phys.
  Rev. Lett.} \textbf{\bibinfo{volume}{113}}, \bibinfo{pages}{205701}
  (\bibinfo{year}{2014}),
  \urlprefix\url{https://link.aps.org/doi/10.1103/PhysRevLett.113.205701}.

\bibitem[{\citenamefont{Heyl}(2015)}]{Heyl2015}
\bibinfo{author}{\bibfnamefont{M.}~\bibnamefont{Heyl}}, \bibinfo{journal}{Phys.
  Rev. Lett.} \textbf{\bibinfo{volume}{115}}, \bibinfo{pages}{140602}
  (\bibinfo{year}{2015}),
  \urlprefix\url{https://link.aps.org/doi/10.1103/PhysRevLett.115.140602}.

\bibitem[{\citenamefont{Andraschko and Sirker}(2014)}]{Andraschko2014}
\bibinfo{author}{\bibfnamefont{F.}~\bibnamefont{Andraschko}} \bibnamefont{and}
  \bibinfo{author}{\bibfnamefont{J.}~\bibnamefont{Sirker}},
  \bibinfo{journal}{Phys. Rev. B} \textbf{\bibinfo{volume}{89}},
  \bibinfo{pages}{125120} (\bibinfo{year}{2014}),
  \urlprefix\url{https://link.aps.org/doi/10.1103/PhysRevB.89.125120}.

\bibitem[{\citenamefont{Vajna and D\'ora}(2014)}]{Vajna2014}
\bibinfo{author}{\bibfnamefont{S.}~\bibnamefont{Vajna}} \bibnamefont{and}
  \bibinfo{author}{\bibfnamefont{B.}~\bibnamefont{D\'ora}},
  \bibinfo{journal}{Phys. Rev. B} \textbf{\bibinfo{volume}{89}},
  \bibinfo{pages}{161105} (\bibinfo{year}{2014}),
  \urlprefix\url{https://link.aps.org/doi/10.1103/PhysRevB.89.161105}.

\bibitem[{\citenamefont{Budich and Heyl}(2016)}]{Budich2016}
\bibinfo{author}{\bibfnamefont{J.~C.} \bibnamefont{Budich}} \bibnamefont{and}
  \bibinfo{author}{\bibfnamefont{M.}~\bibnamefont{Heyl}},
  \bibinfo{journal}{Phys. Rev. B} \textbf{\bibinfo{volume}{93}},
  \bibinfo{pages}{085416} (\bibinfo{year}{2016}),
  \urlprefix\url{https://link.aps.org/doi/10.1103/PhysRevB.93.085416}.

\bibitem[{\citenamefont{Bhattacharya et~al.}(2017)\citenamefont{Bhattacharya,
  Bandyopadhyay, and Dutta}}]{Bhattacharya2017}
\bibinfo{author}{\bibfnamefont{U.}~\bibnamefont{Bhattacharya}},
  \bibinfo{author}{\bibfnamefont{S.}~\bibnamefont{Bandyopadhyay}},
  \bibnamefont{and} \bibinfo{author}{\bibfnamefont{A.}~\bibnamefont{Dutta}},
  \bibinfo{journal}{Phys. Rev. B} \textbf{\bibinfo{volume}{96}},
  \bibinfo{pages}{180303} (\bibinfo{year}{2017}),
  \urlprefix\url{https://link.aps.org/doi/10.1103/PhysRevB.96.180303}.

\bibitem[{\citenamefont{Heyl and Budich}(2017)}]{Heyl2017}
\bibinfo{author}{\bibfnamefont{M.}~\bibnamefont{Heyl}} \bibnamefont{and}
  \bibinfo{author}{\bibfnamefont{J.~C.} \bibnamefont{Budich}},
  \bibinfo{journal}{Phys. Rev. B} \textbf{\bibinfo{volume}{96}},
  \bibinfo{pages}{180304} (\bibinfo{year}{2017}),
  \urlprefix\url{https://link.aps.org/doi/10.1103/PhysRevB.96.180304}.

\bibitem[{\citenamefont{Jurcevic et~al.}(2017)\citenamefont{Jurcevic, Shen,
  Hauke, Maier, Brydges, Hempel, Lanyon, Heyl, Blatt, and Roos}}]{Jurcevic2017}
\bibinfo{author}{\bibfnamefont{P.}~\bibnamefont{Jurcevic}},
  \bibinfo{author}{\bibfnamefont{H.}~\bibnamefont{Shen}},
  \bibinfo{author}{\bibfnamefont{P.}~\bibnamefont{Hauke}},
  \bibinfo{author}{\bibfnamefont{C.}~\bibnamefont{Maier}},
  \bibinfo{author}{\bibfnamefont{T.}~\bibnamefont{Brydges}},
  \bibinfo{author}{\bibfnamefont{C.}~\bibnamefont{Hempel}},
  \bibinfo{author}{\bibfnamefont{B.~P.} \bibnamefont{Lanyon}},
  \bibinfo{author}{\bibfnamefont{M.}~\bibnamefont{Heyl}},
  \bibinfo{author}{\bibfnamefont{R.}~\bibnamefont{Blatt}}, \bibnamefont{and}
  \bibinfo{author}{\bibfnamefont{C.~F.} \bibnamefont{Roos}},
  \bibinfo{journal}{Phys. Rev. Lett.} \textbf{\bibinfo{volume}{119}},
  \bibinfo{pages}{080501} (\bibinfo{year}{2017}),
  \urlprefix\url{https://link.aps.org/doi/10.1103/PhysRevLett.119.080501}.

\bibitem[{\citenamefont{Fl{\"a}schner et~al.}(2018)\citenamefont{Fl{\"a}schner,
  Vogel, Tarnowski, Rem, L{\"u}hmann, Heyl, Budich, Mathey, Sengstock, and
  Weitenberg}}]{Flaeschner2018}
\bibinfo{author}{\bibfnamefont{N.}~\bibnamefont{Fl{\"a}schner}},
  \bibinfo{author}{\bibfnamefont{D.}~\bibnamefont{Vogel}},
  \bibinfo{author}{\bibfnamefont{M.}~\bibnamefont{Tarnowski}},
  \bibinfo{author}{\bibfnamefont{B.~S.} \bibnamefont{Rem}},
  \bibinfo{author}{\bibfnamefont{D.-S.} \bibnamefont{L{\"u}hmann}},
  \bibinfo{author}{\bibfnamefont{M.}~\bibnamefont{Heyl}},
  \bibinfo{author}{\bibfnamefont{J.~C.} \bibnamefont{Budich}},
  \bibinfo{author}{\bibfnamefont{L.}~\bibnamefont{Mathey}},
  \bibinfo{author}{\bibfnamefont{K.}~\bibnamefont{Sengstock}},
  \bibnamefont{and}
  \bibinfo{author}{\bibfnamefont{C.}~\bibnamefont{Weitenberg}},
  \bibinfo{journal}{Nature Physics} \textbf{\bibinfo{volume}{14}},
  \bibinfo{pages}{265} (\bibinfo{year}{2018}), ISSN \bibinfo{issn}{1745-2481},
  \urlprefix\url{https://doi.org/10.1038/s41567-017-0013-8}.

\bibitem[{\citenamefont{Calabrese et~al.}(2011)\citenamefont{Calabrese, Essler,
  and Fagotti}}]{Calabrese2011}
\bibinfo{author}{\bibfnamefont{P.}~\bibnamefont{Calabrese}},
  \bibinfo{author}{\bibfnamefont{F.~H.~L.} \bibnamefont{Essler}},
  \bibnamefont{and} \bibinfo{author}{\bibfnamefont{M.}~\bibnamefont{Fagotti}},
  \bibinfo{journal}{Phys. Rev. Lett.} \textbf{\bibinfo{volume}{106}},
  \bibinfo{pages}{227203} (\bibinfo{year}{2011}),
  \urlprefix\url{https://link.aps.org/doi/10.1103/PhysRevLett.106.227203}.

\bibitem[{\citenamefont{Calabrese et~al.}(2012)\citenamefont{Calabrese, Essler,
  and Fagotti}}]{Calabrese2012}
\bibinfo{author}{\bibfnamefont{P.}~\bibnamefont{Calabrese}},
  \bibinfo{author}{\bibfnamefont{F.~H.~L.} \bibnamefont{Essler}},
  \bibnamefont{and} \bibinfo{author}{\bibfnamefont{M.}~\bibnamefont{Fagotti}},
  \bibinfo{journal}{Journal of Statistical Mechanics: Theory and Experiment}
  \textbf{\bibinfo{volume}{2012}}, \bibinfo{pages}{P07016}
  (\bibinfo{year}{2012}),
  \urlprefix\url{http://stacks.iop.org/1742-5468/2012/i=07/a=P07016}.

\bibitem[{\citenamefont{Schmitt and Kehrein}(2015)}]{Schmitt2015}
\bibinfo{author}{\bibfnamefont{M.}~\bibnamefont{Schmitt}} \bibnamefont{and}
  \bibinfo{author}{\bibfnamefont{S.}~\bibnamefont{Kehrein}},
  \bibinfo{journal}{Phys. Rev. B} \textbf{\bibinfo{volume}{92}},
  \bibinfo{pages}{075114} (\bibinfo{year}{2015}),
  \urlprefix\url{https://link.aps.org/doi/10.1103/PhysRevB.92.075114}.

\bibitem[{\citenamefont{Bhattacharya and
  Dutta}(2017{\natexlab{a}})}]{Bhattacharya2017b}
\bibinfo{author}{\bibfnamefont{U.}~\bibnamefont{Bhattacharya}}
  \bibnamefont{and} \bibinfo{author}{\bibfnamefont{A.}~\bibnamefont{Dutta}},
  \bibinfo{journal}{Phys. Rev. B} \textbf{\bibinfo{volume}{96}},
  \bibinfo{pages}{014302} (\bibinfo{year}{2017}{\natexlab{a}}),
  \urlprefix\url{https://link.aps.org/doi/10.1103/PhysRevB.96.014302}.

\bibitem[{\citenamefont{Bhattacharya and
  Dutta}(2017{\natexlab{b}})}]{Bhattacharya2017c}
\bibinfo{author}{\bibfnamefont{U.}~\bibnamefont{Bhattacharya}}
  \bibnamefont{and} \bibinfo{author}{\bibfnamefont{A.}~\bibnamefont{Dutta}},
  \bibinfo{journal}{Phys. Rev. B} \textbf{\bibinfo{volume}{95}},
  \bibinfo{pages}{184307} (\bibinfo{year}{2017}{\natexlab{b}}),
  \urlprefix\url{https://link.aps.org/doi/10.1103/PhysRevB.95.184307}.

\bibitem[{\citenamefont{Weidinger et~al.}(2017)\citenamefont{Weidinger, Heyl,
  Silva, and Knap}}]{Weidinger2017}
\bibinfo{author}{\bibfnamefont{S.~A.} \bibnamefont{Weidinger}},
  \bibinfo{author}{\bibfnamefont{M.}~\bibnamefont{Heyl}},
  \bibinfo{author}{\bibfnamefont{A.}~\bibnamefont{Silva}}, \bibnamefont{and}
  \bibinfo{author}{\bibfnamefont{M.}~\bibnamefont{Knap}},
  \bibinfo{journal}{Phys. Rev. B} \textbf{\bibinfo{volume}{96}},
  \bibinfo{pages}{134313} (\bibinfo{year}{2017}),
  \urlprefix\url{https://link.aps.org/doi/10.1103/PhysRevB.96.134313}.

\bibitem[{\citenamefont{{Heyl} et~al.}(2018)\citenamefont{{Heyl}, {Pollmann},
  and {D\'ora}}}]{Heyl2018}
\bibinfo{author}{\bibfnamefont{M.}~\bibnamefont{{Heyl}}},
  \bibinfo{author}{\bibfnamefont{F.}~\bibnamefont{{Pollmann}}},
  \bibnamefont{and} \bibinfo{author}{\bibfnamefont{B.}~\bibnamefont{{D\'ora}}},
  \bibinfo{journal}{ArXiv e-prints}  (\bibinfo{year}{2018}),
  \eprint{1801.01684}, \urlprefix\url{https://arxiv.org/abs/1801.01684}.

\bibitem[{\citenamefont{{De Nicola} et~al.}(2018)\citenamefont{{De Nicola},
  {Doyon}, and {Bhaseen}}}]{DeNicola2018}
\bibinfo{author}{\bibfnamefont{S.}~\bibnamefont{{De Nicola}}},
  \bibinfo{author}{\bibfnamefont{B.}~\bibnamefont{{Doyon}}}, \bibnamefont{and}
  \bibinfo{author}{\bibfnamefont{M.~J.} \bibnamefont{{Bhaseen}}},
  \bibinfo{journal}{ArXiv e-prints}  (\bibinfo{year}{2018}),
  \eprint{1805.05350}, \urlprefix\url{https://arxiv.org/abs/1805.05350}.

\bibitem[{\citenamefont{Zauner-Stauber and Halimeh}(2017)}]{Zauner2017}
\bibinfo{author}{\bibfnamefont{V.}~\bibnamefont{Zauner-Stauber}}
  \bibnamefont{and} \bibinfo{author}{\bibfnamefont{J.~C.}
  \bibnamefont{Halimeh}}, \bibinfo{journal}{Phys. Rev. E}
  \textbf{\bibinfo{volume}{96}}, \bibinfo{pages}{062118}
  (\bibinfo{year}{2017}),
  \urlprefix\url{https://link.aps.org/doi/10.1103/PhysRevE.96.062118}.

\bibitem[{\citenamefont{{Halimeh} et~al.}(2018)\citenamefont{{Halimeh}, {Van
  Damme}, {Zauner-Stauber}, and {Vanderstraeten}}}]{Halimeh2018}
\bibinfo{author}{\bibfnamefont{J.~C.} \bibnamefont{{Halimeh}}},
  \bibinfo{author}{\bibfnamefont{M.}~\bibnamefont{{Van Damme}}},
  \bibinfo{author}{\bibfnamefont{V.}~\bibnamefont{{Zauner-Stauber}}},
  \bibnamefont{and}
  \bibinfo{author}{\bibfnamefont{L.}~\bibnamefont{{Vanderstraeten}}},
  \bibinfo{journal}{ArXiv e-prints}  (\bibinfo{year}{2018}),
  \eprint{1810.07187}, \urlprefix\url{https://arxiv.org/abs/1810.07187}.

\bibitem[{\citenamefont{Defenu et~al.}(2019)\citenamefont{Defenu, Enss, and
  Halimeh}}]{Defenu2019}
\bibinfo{author}{\bibfnamefont{N.}~\bibnamefont{Defenu}},
  \bibinfo{author}{\bibfnamefont{T.}~\bibnamefont{Enss}}, \bibnamefont{and}
  \bibinfo{author}{\bibfnamefont{J.~C.} \bibnamefont{Halimeh}},
  \bibinfo{journal}{Phys. Rev. B} \textbf{\bibinfo{volume}{100}},
  \bibinfo{pages}{014434} (\bibinfo{year}{2019}),
  \urlprefix\url{https://link.aps.org/doi/10.1103/PhysRevB.100.014434}.

\bibitem[{\citenamefont{du~Croo~de Jongh and van
  Leeuwen}(1998)}]{duCroodeJongh1998}
\bibinfo{author}{\bibfnamefont{M.~S.~L.} \bibnamefont{du~Croo~de Jongh}}
  \bibnamefont{and} \bibinfo{author}{\bibfnamefont{J.~M.~J.} \bibnamefont{van
  Leeuwen}}, \bibinfo{journal}{Phys. Rev. B} \textbf{\bibinfo{volume}{57}},
  \bibinfo{pages}{8494} (\bibinfo{year}{1998}),
  \urlprefix\url{https://link.aps.org/doi/10.1103/PhysRevB.57.8494}.

\bibitem[{\citenamefont{Bl\"ote and Deng}(2002)}]{Bloete2002}
\bibinfo{author}{\bibfnamefont{H.~W.~J.} \bibnamefont{Bl\"ote}}
  \bibnamefont{and} \bibinfo{author}{\bibfnamefont{Y.}~\bibnamefont{Deng}},
  \bibinfo{journal}{Phys. Rev. E} \textbf{\bibinfo{volume}{66}},
  \bibinfo{pages}{066110} (\bibinfo{year}{2002}),
  \urlprefix\url{https://link.aps.org/doi/10.1103/PhysRevE.66.066110}.

\bibitem[{\citenamefont{{McCulloch}}(2008)}]{McCulloch2008}
\bibinfo{author}{\bibfnamefont{I.~P.} \bibnamefont{{McCulloch}}},
  \bibinfo{journal}{ArXiv e-prints}  (\bibinfo{year}{2008}),
  \eprint{0804.2509}, \urlprefix\url{https://arxiv.org/abs/0804.2509}.

\bibitem[{Ian()}]{IanToolkit}
\bibinfo{howpublished}{See
  https://people.smp.uq.edu.au/IanMcCulloch/mptoolkit/.}

\bibitem[{mpt()}]{mptoolkit}
\bibinfo{howpublished}{Ian P. McCulloch, \textit{Matrix Product Toolkit}. URL:
  http://physics.uq.edu.au/people/ianmcc/mptoolkit/}.

\bibitem[{\citenamefont{Hashizume et~al.}(2020)\citenamefont{Hashizume,
  Halimeh, and McCulloch}}]{Hashizume2019}
\bibinfo{author}{\bibfnamefont{T.}~\bibnamefont{Hashizume}},
  \bibinfo{author}{\bibfnamefont{J.~C.} \bibnamefont{Halimeh}},
  \bibnamefont{and} \bibinfo{author}{\bibfnamefont{I.~P.}
  \bibnamefont{McCulloch}}, \bibinfo{journal}{Phys. Rev. B}
  \textbf{\bibinfo{volume}{102}}, \bibinfo{pages}{035115}
  (\bibinfo{year}{2020}),
  \urlprefix\url{https://link.aps.org/doi/10.1103/PhysRevB.102.035115}.

\bibitem[{\citenamefont{{Liu} et~al.}(2018)\citenamefont{{Liu}, {Lundgren},
  {Titum}, {Pagano}, {Zhang}, {Monroe}, and {Gorshkov}}}]{Liu2018}
\bibinfo{author}{\bibfnamefont{F.}~\bibnamefont{{Liu}}},
  \bibinfo{author}{\bibfnamefont{R.}~\bibnamefont{{Lundgren}}},
  \bibinfo{author}{\bibfnamefont{P.}~\bibnamefont{{Titum}}},
  \bibinfo{author}{\bibfnamefont{G.}~\bibnamefont{{Pagano}}},
  \bibinfo{author}{\bibfnamefont{J.}~\bibnamefont{{Zhang}}},
  \bibinfo{author}{\bibfnamefont{C.}~\bibnamefont{{Monroe}}}, \bibnamefont{and}
  \bibinfo{author}{\bibfnamefont{A.~V.} \bibnamefont{{Gorshkov}}},
  \bibinfo{journal}{ArXiv e-prints}  (\bibinfo{year}{2018}),
  \eprint{1810.02365}, \urlprefix\url{https://arxiv.org/abs/1810.02365}.

\bibitem[{\citenamefont{Zeiher et~al.}(2016)\citenamefont{Zeiher, van Bijnen,
  Schau\ss, Hild, Choi, Pohl, Bloch, and Gross}}]{Zeiher2016}
\bibinfo{author}{\bibfnamefont{J.}~\bibnamefont{Zeiher}},
  \bibinfo{author}{\bibfnamefont{R.}~\bibnamefont{van Bijnen}},
  \bibinfo{author}{\bibfnamefont{P.}~\bibnamefont{Schau\ss}},
  \bibinfo{author}{\bibfnamefont{S.}~\bibnamefont{Hild}},
  \bibinfo{author}{\bibfnamefont{J.-y.} \bibnamefont{Choi}},
  \bibinfo{author}{\bibfnamefont{P.}~\bibnamefont{Pohl}},
  \bibinfo{author}{\bibfnamefont{I.}~\bibnamefont{Bloch}}, \bibnamefont{and}
  \bibinfo{author}{\bibfnamefont{C.}~\bibnamefont{Gross}},
  \bibinfo{journal}{Nature Physics} \textbf{\bibinfo{volume}{12}},
  \bibinfo{pages}{1095} (\bibinfo{year}{2016}),
  \urlprefix\url{https://www.nature.com/articles/nphys3835}.

\bibitem[{\citenamefont{Gross and Bloch}(2017)}]{Gross2017}
\bibinfo{author}{\bibfnamefont{C.}~\bibnamefont{Gross}} \bibnamefont{and}
  \bibinfo{author}{\bibfnamefont{I.}~\bibnamefont{Bloch}},
  \bibinfo{journal}{Science} \textbf{\bibinfo{volume}{357}},
  \bibinfo{pages}{995} (\bibinfo{year}{2017}), ISSN \bibinfo{issn}{0036-8075},
  \urlprefix\url{http://science.sciencemag.org/content/357/6355/995}.

\bibitem[{\citenamefont{Uhrich et~al.}(2020)\citenamefont{Uhrich, Defenu,
  Jafari, and Halimeh}}]{Uhrich2020}
\bibinfo{author}{\bibfnamefont{P.}~\bibnamefont{Uhrich}},
  \bibinfo{author}{\bibfnamefont{N.}~\bibnamefont{Defenu}},
  \bibinfo{author}{\bibfnamefont{R.}~\bibnamefont{Jafari}}, \bibnamefont{and}
  \bibinfo{author}{\bibfnamefont{J.~C.} \bibnamefont{Halimeh}},
  \bibinfo{journal}{Phys. Rev. B} \textbf{\bibinfo{volume}{101}},
  \bibinfo{pages}{245148} (\bibinfo{year}{2020}),
  \urlprefix\url{https://link.aps.org/doi/10.1103/PhysRevB.101.245148}.

\bibitem[{\citenamefont{Trotter}(1959)}]{Trotter1959}
\bibinfo{author}{\bibfnamefont{H.~F.} \bibnamefont{Trotter}},
  \bibinfo{journal}{Proc. Amer. Math. Soc.} \textbf{\bibinfo{volume}{10}},
  \bibinfo{pages}{545} (\bibinfo{year}{1959}),
  \urlprefix\url{https://doi.org/10.1090/S0002-9939-1959-0108732-6}.

\bibitem[{\citenamefont{Suzuki}(1976)}]{Suzuki1976}
\bibinfo{author}{\bibfnamefont{M.}~\bibnamefont{Suzuki}},
  \bibinfo{journal}{Progress of Theoretical Physics}
  \textbf{\bibinfo{volume}{56}}, \bibinfo{pages}{1454} (\bibinfo{year}{1976}),
  \urlprefix\url{https://dx.doi.org/10.1143/PTP.56.1454}.

\bibitem[{\citenamefont{Noack and Manmana}(2005)}]{Noack2005}
\bibinfo{author}{\bibfnamefont{R.~M.} \bibnamefont{Noack}} \bibnamefont{and}
  \bibinfo{author}{\bibfnamefont{S.~R.} \bibnamefont{Manmana}},
  \emph{\bibinfo{title}{{Diagonalization- and numerical
  renormalization-group-based methods for interacting quantum systems}}}, vol.
  \bibinfo{volume}{789} (\bibinfo{year}{2005}), ISBN
  \bibinfo{isbn}{0735402795}, \eprint{0510321}.

\bibitem[{\citenamefont{Hochbruck and Lubich}(1997)}]{Mathematics2015}
\bibinfo{author}{\bibfnamefont{M.}~\bibnamefont{Hochbruck}} \bibnamefont{and}
  \bibinfo{author}{\bibfnamefont{C.}~\bibnamefont{Lubich}},
  \bibinfo{journal}{SIAM Journal on Numerical Analysis}
  \textbf{\bibinfo{volume}{34}}, \bibinfo{pages}{1911} (\bibinfo{year}{1997}),
  ISSN \bibinfo{issn}{0036-1429},
  \urlprefix\url{http://epubs.siam.org/doi/10.1137/S0036142995280572}.

\bibitem[{\citenamefont{Garc{\'{i}}a-Ripoll}(2006)}]{GarcIa-Ripoll2006}
\bibinfo{author}{\bibfnamefont{J.~J.} \bibnamefont{Garc{\'{i}}a-Ripoll}},
  \bibinfo{journal}{New Journal of Physics} \textbf{\bibinfo{volume}{8}},
  \bibinfo{pages}{305} (\bibinfo{year}{2006}), ISSN \bibinfo{issn}{1367-2630},
  \eprint{0610210},
  \urlprefix\url{http://stacks.iop.org/1367-2630/8/i=12/a=305?key=crossref.1ba36a6744b1adce73f0ed1217be040e}.

\bibitem[{\citenamefont{Or{\'{u}}s and Vidal}(2008)}]{Orus2008}
\bibinfo{author}{\bibfnamefont{R.}~\bibnamefont{Or{\'{u}}s}} \bibnamefont{and}
  \bibinfo{author}{\bibfnamefont{G.}~\bibnamefont{Vidal}},
  \bibinfo{journal}{Physical Review B} \textbf{\bibinfo{volume}{78}},
  \bibinfo{pages}{155117} (\bibinfo{year}{2008}), ISSN
  \bibinfo{issn}{1098-0121},
  \urlprefix\url{https://link.aps.org/doi/10.1103/PhysRevB.78.155117}.

\bibitem[{\citenamefont{Vidal}(2004)}]{Vidal2004}
\bibinfo{author}{\bibfnamefont{G.}~\bibnamefont{Vidal}},
  \bibinfo{journal}{Physical Review Letters} \textbf{\bibinfo{volume}{93}},
  \bibinfo{pages}{040502} (\bibinfo{year}{2004}), ISSN
  \bibinfo{issn}{0031-9007}, \eprint{0310089},
  \urlprefix\url{https://link.aps.org/doi/10.1103/PhysRevLett.93.040502}.

\bibitem[{Has()}]{Hashizume_thesis}
\bibinfo{howpublished}{Tomohiro Hashizume, Honours Thesis, University of
  Queensland, 2017}.

\bibitem[{\citenamefont{Phien et~al.}(2012)\citenamefont{Phien, Vidal, and
  McCulloch}}]{Phien2012}
\bibinfo{author}{\bibfnamefont{H.~N.} \bibnamefont{Phien}},
  \bibinfo{author}{\bibfnamefont{G.}~\bibnamefont{Vidal}}, \bibnamefont{and}
  \bibinfo{author}{\bibfnamefont{I.~P.} \bibnamefont{McCulloch}},
  \bibinfo{journal}{Phys. Rev. B} \textbf{\bibinfo{volume}{86}},
  \bibinfo{pages}{245107} (\bibinfo{year}{2012}),
  \urlprefix\url{https://link.aps.org/doi/10.1103/PhysRevB.86.245107}.

\end{thebibliography}
\end{document}